\documentclass[fleqn,usenatbib]{mnras}

\usepackage{mathptmx}
\usepackage{txfonts}

\usepackage[T1]{fontenc}
\usepackage{ae,aecompl}

\usepackage{graphicx}	
\usepackage{amssymb}	
\usepackage{times}
\usepackage{array}
\usepackage{stfloats}
\usepackage{xspace}
\usepackage{ulem}                   
\usepackage{hyperref}
\usepackage[dvipsnames]{xcolor}

\newcommand{\music}{\textsc{MUSIC}\xspace}

\newcommand{\hMsol}{h^{-1}\,{\rm M_\odot}}

\newcommand{\hMpc}{h^{-1}\,{\rm Mpc}}
\newcommand{\hkpc}{h^{-1}\,{\rm kpc}}
\newcommand{\kms}{{\rm km\, s^{-1}}}

\newcommand{\Lbox}{L_\mathrm{box}}
\newcommand{\HI}{HI\xspace}
\newcommand{\HeI}{HeI\xspace}
\newcommand{\HeII}{HeII\xspace}

\newcommand{\arepo}{{\sc arepo}\xspace}
\newcommand{\areport}{{\sc arepo-RT}\xspace}

\newcommand{\Mtm}{M_\mathrm{200m}}
\newcommand{\Rhms}{R_\mathrm{hmr,*}}
\newcommand{\Mhms}{M_\mathrm{hmr,*}}

\newcommand{\aunone}     {\textsc{Au-none}\xspace}
\newcommand{\aucosmo}    {\textsc{Au-cosmo}\xspace}
\newcommand{\aubier}     {\textsc{Au-Bier}\xspace}
\newcommand{\audurr}     {\textsc{Au-Durr}\xspace}
\newcommand{\audurrlt}   {\textsc{Au-Durr-L3}\xspace}
\newcommand{\ausneH}     {\textsc{Au-SNe-H}\xspace}
\newcommand{\ausne}      {\textsc{Au-SNe}\xspace}

\newcommand{\fbcosmo}    {\textsc{FB-cosmo}\xspace}
\newcommand{\fbcosmonort}{\textsc{FB-cosmo-noRT}\xspace}
\newcommand{\fbbier}     {\textsc{FB-Bier}\xspace}
\newcommand{\fbdurr}     {\textsc{FB-Durr}\xspace}
\newcommand{\fbsneH}     {\textsc{FB-SNe-H}\xspace}
\newcommand{\fbsne}      {\textsc{FB-SNe}\xspace}

\newcommand{\eg}{e.g.\xspace}
\newcommand{\ie}{i.e.\xspace}

\newcommand{\citenp}[1]{\citeauthor{#1} \citeyear{#1}}

\newcommand{\enrico}[1]{#1}%\textcolor{Green}{\textbf{#1}}}

\title[Different magnetic field seeds in galaxy formation]{Magnetogenesis around the first galaxies: the impact of different field seeding processes on galaxy formation}

\author[E. Garaldi et al.]{
Enrico Garaldi,$^{1}$\thanks{E-mail: egaraldi@mpa-garching.mpg.de} R\"udiger Pakmor,$^{1}$ Volker Springel$^{1}$
\\
$^{1}$Max-Planck-Institut f\"ur Astrophysik, Karl-Schwarzschild-Str. 1, Garching b. M\"unchen 85741, Germany\\
}

\date{Accepted XXX. Received YYY; in original form ZZZ}

\pubyear{2020}

\begin{document}
\label{firstpage}
\pagerange{\pageref{firstpage}--\pageref{lastpage}}
\maketitle

\begin{abstract}
We study the evolution of magnetic fields generated by charge segregation ahead of ionization fronts during the Epoch of Reionization, and their effects on galaxy formation. We compare this magnetic seeding process with the Biermann battery, injection from supernovae, and an imposed seed field at redshift $z\gtrsim127$. Using a suite of self-consistent cosmological and zoom-in simulations based on the Auriga galaxy-formation model, we determine that all mechanisms produce galactic magnetic fields that equally affect galaxy formation, and are nearly indistinguishable at $z\lesssim1.5$. The former is compatible with observed values, while the latter is correlated with the gas metallicity below a seed-dependent redshift. Low-density gas and haloes below a seed-dependent mass threshold retain memory of the initial magnetic field. We produce synthetic Faraday rotation measure maps, showing that they have the potential to constrain the seeding process, although current observations are not yet sensitive enough. Our results imply that the ad-hoc assumption of a primordial seed field --~widely used in galaxy formation simulations but of uncertain physical origin~-- can be replaced by physically-motivated mechanisms for magnetogenesis with negligible impact on galactic properties. Additionally, magnetic fields generated ahead of ionization fronts appear very similar but weaker than those produced by the Biermann battery. Hence, in a realistic scenario where both mechanisms are active, the former will be negligible compared to the latter. Finally, our results highlight that the high-redshift Universe is a fruitful testing ground for our understanding of magnetic fields generation.
\end{abstract}


\section{Introduction}
\label{sec:introduction}
Cosmic magnetic fields have been observed at many scales, from planetary and stellar bodies to supra-galactic scales \citep[\eg][]{Beck&Wielebinski2013}. They are not only passive by-products of astrophysical and cosmological processes, but play an important role in many phenomena, \eg accretion on compact objects, propagation of relativistic charged particles, and for the pressure support against gravitational collapse in the inter-stellar medium (ISM).

Despite their ubiquity and relevance, the origin of astrophysical magnetic fields remains a mystery. Over the course of many decades, a number of different mechanisms able to generate initial small \textit{seed} fields have been proposed, but no conclusive evidence could be gathered on which of them is dominant. There are a number of reasons for this situation. Firstly, observing magnetic fields is very challenging, since it involves indirect measurements that often are plagued by significant degeneracies with other physical quantities \citep{Beck+15}. A second reason is that most of the available constraints come from galaxies \citep[\eg][]{Tullmann+2000, Mao+12, Beck12, Beck+15, Damas-Segovia+16, Terral&Ferriere17, Stein+19}, galaxy clusters \citep[\eg][]{Kim+91, Feretti+99, Clarke+01, Johnston-Hollitt+04, Govoni+06, Guidetti+08, Bonafede+10, vanWeeren+19review} or --~more recently~-- the circum-galactic medium \citep[CGM][]{Bernet+2008, Prochaska+19}. These are regions where the magnetic field has been heavily re-processed by physical processes in the galactic disk, and hence may have completely lost memory of its origin. \enrico{There are, however, attempts to constrain the field strength in the cosmic web and in the inter-galactic medium (IGM), mostly resulting in upper limits \citep[see \eg ][]{Brown+17, Vernstrom+17, Vernstrom+19, OSullivan+20}}.
Finally, most constraints come from redshift $z\lesssim1$, rendering it hard to disentangle the original seed field from the transformation and amplification it underwent throughout billions of years.

Magnetic field seeding mechanisms can be broadly categorized into two classes. The first comprises those that generate magnetic fields on the largest scales, which are then brought into galaxies by gas accretion and amplified there by shear flows \citep{Dolag+99}, or turbulent \citep[\eg][]{Dolag+99, Arshakian+09, Federrath+11, Schleicher+13, Sur+10, Beresnyak&Miniati16, Rieder&Teyssier16, Rieder&Teyssier17a, Pakmor+2017, Federrath+14} and galactic \citep{Hanasz+04, Rieder&Teyssier16, Rieder&Teyssier17a, Pakmor+2017} dynamo processes. All processes at play during inflation, phase transitions in the very early Universe, and large-scale plasma phenomena  fall into this class. An example of the latter is the so-called Biermann battery \citep{Biermann1950}, which creates magnetic fields from the mis-alignments between density and temperature gradients, for instance in cosmological shocks \citep{Ryu+98} and during the Epoch of Reionization \citep[EoR, ][]{Gnedin+2000}.

The second class of seeding mechanisms ties the origin of magnetic fields to stars \citep{Pudritz+89} or AGNs \citep{Daly&Loeb90, Furlanetto&Loeb01} within (proto-)galaxies. These seed fields are then amplified and subsequently dispersed in the CGM and IGM through feedback processes \citep[\eg][]{Donnert+09}. 

Regardless of the seeding process, in ideal magneto-hydrodynamics (MHD) the conservation of magnetic flux  enhances (suppresses) the magnetic field strength $|\mathbf{B}|$ during gas compression (expansion). In absence of any amplification mechanism, $|\mathbf{B}| \propto \rho_\mathrm{b}^{2/3}$, where $\rho_\mathrm{b}$ is the gas density.

Numerical simulations can be of substantial help in the quest to better understand cosmic magnetism. Until recently, most of the effort was concentrated on simulating individual objects (or part of them), in order to reach a resolution sufficiently high to capture amplification processes \citep{Hanasz+04, Rieder&Teyssier16, Rieder&Teyssier17a, Pakmor+2017}. However, thanks to the advances in computational power and simulation algorithms, cosmological simulations which cover large portions of the Universe and which can study magnetic fields in entire galaxy populations have now become possible. Notable examples are the MAGNETICUM project\footnote{\url{http://www.magneticum.org/}}, the IllustrisTNG suite \citep{TNG-Marinacci, TNG-Naiman, TNG-Nelson, TNG-Pillepich, TNG-Springel}, and the simulations presented in \citet{AlvesBatista+2017}, \enrico{ \citet{Hutschenreuter+18}, and \citet[][although in a rather limited volume]{Katz+19} }. Nevertheless, systematic numerical investigations of the different seeding mechanisms are rare. One example is the simulation suite presented in \citet{Vazza2017}, where a variety of mechanisms for magnetogenesis was explored, including both primordial and astrophysical origins, using numerical simulation on a large uniform Cartesian grid. A similar --~although smaller~-- investigation was carried out in \citet{Marinacci+2015} and \citet{Marinacci&Vogelsberger2016}, where the impact of the primordial seed strength on the properties of galaxies has been studied in the context of a uniform seeding at $z \gtrsim 100$, and in \citet{Vazza+20}, where the effects of the seed field power spectrum were investigated.

In this paper, we explore the properties of a new seeding mechanism, which has been recently proposed on the basis of theoretical arguments \citep{Durrive&Langer2015, Durrive+2017} but never explored in a realistic galaxy formation context. In this scenario, magnetic fields are generated by anisotropic ionisation fronts sweeping through the Universe during the EoR (see Sec.~\ref{sec:Durrive_battery} for details). 

In order to provide a holistic view of the properties of this new magnetogenesis mechanism that we dub, for short, `Durrive battery' in analogy with the well-known Biermann battery, we compare it with commonly-employed seeding prescriptions, namely: a uniform seed field of pre-recombination origin (dubbed \textit{cosmological} for simplicity), the Biermann battery (active in the diffuse gas only, we do not include any sub-resolution model for stellar magnetic fields), and a simple magnetic injection scheme related to supernova explosions. We describe these mechanisms as well as their implementation in our simulations  in more depth in Sec.~\ref{sec:magnetic_seed}.  Since even in modern simulations the amplification of existing magnetic fields is still limited by numerical resolution, we take a dual approach and explore magnetic seeding  both in zoom-in and initially uniformly sampled simulations. The former allow us to explore mechanism at play in single galaxies at high resolution, while the latter enables a large-scale view despite having stronger resolution limitations. The technical details of our simulations are described in Sec.~\ref{sec:methods}, while their results are laid out in Sec.~\ref{sec:galactic_magnetic_fields} and Sec.~\ref{sec:large_scales_magnetic_fields}, respectively. Finally, we provide a discussion of the implications of our results in Sec.~\ref{sec:discussion} and concluding remarks in Sec.~\ref{sec:conclusions}.

\section{Magnetic seed field creation}
\label{sec:magnetic_seed}
The main objective of this Paper is to compare four different magnetic seeding mechanisms.  
In the following, we briefly review each of them and discuss our corresponding implementations in the simulation code \arepo (see Sec.~\ref{sec:methods} for an in-depth description of the numerical details of the code).

\subsection{Cosmological seed field}
From a galaxy formation perspective, the simplest way to magnetise the Universe is to assume that a seed field was generated in the very early Universe ($z \gg 100$). There are several mechanisms proposed to achieve this, many of which are linked to the inflation era \citep[for a review, see \eg][]{magnetic_seed_review, Subramanian16review, Kandus+11review}. In the context of this paper, the details of the physical process are not important, as long as the resulting seed field strength is spatially homogeneous on the scale of the simulation box. 

In practice, we implement a cosmological seed by imposing a small uniform magnetic field in the initial conditions of the simulation. In order to maintain a minimum set of assumptions, we choose the simplest field  geometry compatible with the vanishing divergence constraint, \ie field lines aligned with the  $z$-direction of the box, which fold onto themselves thanks to the periodic boundary conditions of the simulation. \citep[More complicated initial field geometries have been recently investigated \eg in][]{Vazza+20}. We select an initial comoving field strength equal to $B_\mathrm{cosmo} (z_\mathrm{init}) = 10^{-14} \, \mathrm{G}$ (corresponding to a physical field strength in the initial conditions of $B_\mathrm{physical} = a^{-2} \, B_\mathrm{comoving} = 2 \times 10^{-11} \, \mathrm{G}$ at $z_\mathrm{init} = 127$. 
Previous studies have shown that this choice has no effect on the final galactic magnetic field, as long as it is below a threshold value of $B_\mathrm{comoving} (z_\mathrm{init}) \approx 10^{-9} \, \mathrm{G}$. Above this value, galaxies themselves are affected and the star-formation-rate evolution is suppressed below the observed level \citep{Marinacci&Vogelsberger2016}.

\subsection{Biermann battery}
Magnetic fields can be created by the vorticity-induced relative motion of ions and electrons, a process dubbed Biermann battery \citep{Biermann1950}. From a macroscopic point of view, this is described by:

\begin{equation}
\label{eq:BiermannBattery}
    \frac{\partial \mathbf{B}}{\partial t} = \frac{c}{e n_e^2} \mathbf{\nabla} p_e \times \mathbf{\nabla} n_e ,
\end{equation}
where $n_e$ and $p_e$ are the electron density and pressure, respectively, $c$ is the speed of light, and $e$ the elementary charge. 
It can be seen that magnetic fields are generated only when the gradients of the electron density and pressure are misaligned. This can occur for instance in the IGM during the EoR, when ionisation fronts encounter a cosmic filament. In this configuration, $\mathbf{\nabla} n_e$ is perpendicular to the ionisation front itself, while $\mathbf{\nabla} p_e$ is aligned along the radial filament direction \citep{Gnedin+2000}. In order to self-consistently simulate the Biermann battery, we solve a discretized version of Eq.~(\ref{eq:BiermannBattery}). This depends only on variables explicitly tracked in the code, except for the electron pressure $p_e$, which we compute assuming the electrons follow a perfect gas law, \ie $p_e = n_e k_B T_e$, with $k_B$ being the Boltzmann constant and $T_e$ being the temperature of free electrons, assumed to be in thermal equilibrium with the gas, and $n_e$ follows from the density and ionization state of the local gas.

\subsection{SN and their remnants}
The process of structure formation can be responsible for the generation of magnetic fields. Proto-galactic clouds can show very low levels of magnetization prior to collapse \citep{Wiechen+1998, Birk+2002}, which can be enhanced by gravitational compression during gas collapse into clouds, and subsequently by a combination of turbulent- and $\alpha-\omega$-dynamo due to the (proto)star rotation \citep[\eg][]{Schleicher+13, Sur+10}. Once stars explode into SNe, the amplified field is distributed into the circum- and inter-stellar medium, where it can undergo further amplification during the shock-medium interaction \citep[\eg][]{Xu&Lazarian2017}.

The processes that produce and amplify magnetic fields in SN are not only uncertain, they also occur on very small spatial and temporal scales (order of pc and hundreds-to-thousand years). Hence, they can not yet be self-consistently incorporated in simulations of galaxy formation. Additionally, it remains unclear whether these fields are long-lived enough to reach the surrounding inter-stellar medium, and what strength they have when the latter occurs. For these reasons, we follow here an empirical approach, injecting for each SN a small fraction $f_B$ of its energy into a dipolar magnetic field around it. We add this field in the $64$ gas cells nearest to the stellar particle that originates the SN explosion, and use a randomly-oriented magnetic dipole for the field geometry. Our fiducial value is $f_B = 10^{-4}$, but we also investigate the effect of a ten-times-higher value in Appendix~\ref{app:compare_sn}. \enrico{The main reason for the choice of this value is that, from our tests, values $f_B \geq 10^{-3}$ result in a strong suppression of the star-formation rate density (see Appendix~\ref{app:compare_sn}). In order to minimize potential sources of differences between the runs and better isolate effects purely due to the magnetic seed injection, we decided to employ the largest value of $f_B$ that does not significantly alter the star-formation rate density. Additionally, this produces field strengths of a few $\mu$G close to the SN events, in broad agreement with --~although slightly lower than~-- observed values \citep[\eg][]{Parizot+06}.} A similar approach has been employed by \citet{Beck+13} and is discussed in Sec.~\ref{sec:discussion}.

\subsection{Durrive battery}
\label{sec:Durrive_battery}
A different mechanism able to magnetise gas during the EoR has recently been proposed by \citet{Durrive&Langer2015} and \citet{Durrive+2017}. 
The underlying physical process is rather simple. When an hydrogen atom is photo-ionised, the excess photon momentum is transferred to both the outgoing proton and electron. However, the unequal mass of these two particles induces a systematic charge segregation ahead of the ionisation front, which in turn generates a small but coherent electric field. The latter has a non-vanishing curl whenever the ionisation front is not locally spherical, generating a magnetic field. This has the following macroscopic description:

\begin{equation}
\label{eq:DurriveBattery}
    \frac{\partial \mathbf{B}}{\partial t} = \frac{c}{e n_e^2} \mathbf{\nabla} n_e \times \mathbf{\dot{p}}_\mathrm{pe} - \frac{c}{e n_e} \mathbf{\nabla} \times \mathbf{\dot{p}}_\mathrm{pe},
\end{equation}
where $\mathbf{\dot{p}}_\mathrm{pe}$ is the momentum transferred to the photo-ionised electron. 

The Durrive battery has its roots in \citet{Langer+2003}, where the process was originally presented. In that work, however, a steady-state assumption was used, which turned out not to be applicable since the steady state requires a time larger than the Hubble time to be reached \citep{Ando+2010}. In subsequent works, \citet{Langer+2005} revisited the predictions for the seed strength, and \citet{Doi&Susa2011} showed, using both analytical arguments and idealised simulations, that the photo-generation of magnetic fields is sub-dominant (by approximately $1$ order of magnitude) with respect to the Biermann battery in the same system. However, they focused on large overdensities in the proximity of the photon source, where the latter is most effective. \citet{Durrive&Langer2015} argued that in regions farther away and in milder overdensities, the photo-magnetisation is dominant. Note, however, that all these considerations assume very idealised configurations, which do not capture the complexity of cosmological gas distributions and --~most notably~-- feedback processes.

In practice, we  self-consistently solve Eq.~(\ref{eq:DurriveBattery}) in the code. We extend the work of \citet{Durrive&Langer2015} by including also the contribution coming from helium-ionizing photons. In order to compute $\mathbf{\dot{p}}_\mathrm{pe}$ from the excess energy of ionizing photons, we require the fraction of incoming photon momentum transferred to the electron. We rely on \citet{Sommerfeld&Schurr1930} for the case of hydrogen, on \citet{Wang+2017} for \HeI, and on \citet{Massacrier&El-Murr1996} for \HeII. In Appendix~\ref{app:Durrive_implementation}, we show an example of such a self-consistent calculation.

\section{Simulations}
\label{sec:methods}

In this paper we employ two sets of simulations: a suite of zoom-in runs (identified with the prefix \textsc{Au} followed by the seeding mechanism employed) that concentrate the computational effort on a single galaxy while retaining its cosmological context, and a set of runs where the resolution is (initially) uniform in the entire simulation domain (hence following the evolution of an entire portion of the Universe, identified with the prefix \textsc{FB}, a shorthand for \textit{full box resolution}). Within each suite, we employ identical initial conditions and simulation parameters, with the obvious exception of the magnetic seeding mechanism. 

Both simulation suites employ a setup for the galaxy formation physics very similar to the one used in the Auriga project\footnote{\url{https://wwwmpa.mpa-garching.mpg.de/auriga}} \citep{Auriga}, with the exception of black holes, whose radiation cannot be currently tracked. This choice is dictated by two main reasons, namely: (i) at $z=0$, the model produces realistic galaxies similar to the Milky Way, which provides beyond doubts the best data available; and (ii) we wish to exploit and build upon the large amount of knowledge gathered for this simulation methodology over the years, which provides an exquisite view of the processes at play during galaxy formation. In the following, we briefly summarise the numerical implementation and physical properties of both suites, highlighting the limited differences with respect to the original Auriga model, and otherwise refer the reader to \citet{Auriga} for a detailed description. Note that, in order to be as consistent as possible, we employ the Auriga model also for our cosmological runs, where it has not been applied before. However, we note that the Auriga model differs only in a number of details from the model realized in IllustrisTNG which has proven very successful in cosmological runs.

\subsection{The simulation code}
Our simulations are carried out using the \arepo code \citep{arepo}, which couples an $N$-body solver for gravity with a Voronoi-mesh-based solver for the (ideal) MHD equations. The former combines a particle-mesh approach (based on the fast Fourier Transform) for computing the long-range force with a hierarchical oct-tree for calculating the short-range force. The MHD equations are solved on an unstructured mesh, built from the Voronoi tesselation of a set of mesh-generating points which follow the flow of the simulated gas, providing a natural and continuous way to automatically adapt the resolution based on the density field. The equations themselves are integrated with a second-order Runge-Kutta scheme, where spatial gradients are estimated from the primitive variables using least-square fitting \citep{Pakmor+2016}. We employ the divergence-cleaning scheme of \citet{Powell+1999} to ensure $\nabla \cdot \mathbf{B} \approx 0$ \citep[for a study of its effectiveness in an essentially-identical setup, see ][]{Pakmor&Springel2013, Pakmor+2020}.

As discussed in Sec.~\ref{sec:magnetic_seed}, some of our magnetic seeding mechanisms require that we self-consistently follow  radiation transport (RT). For consistency, we include the latter in all simulations by employing the \areport \citep{arepo-rt} extension of the \arepo code, which solves the first two moments of the RT equation, paired with an M1 closure relation \citep{M1-Levermore}. Photons are injected by stellar particles into their neighbouring cells in an isotropic way, with a mass-, age- and metallicity-dependent flux computed using the \citet{BC03} stellar population libraries and assuming a \citet{Chabrier03} initial mass function for the stars. \areport solves the full non-equilibrium thermo-chemistry of hydrogen and helium to compute the gas ionisation state and net cooling rate. The latter includes gas cooling via primordial and metal-line processes, self-shielding corrections, and photo-heating. We follow radiation with energy ranging from $13.6$ eV to $100$ eV, discretised in frequency space into three bands covering, respectively, the ionisation energies of \HI, \HeI, and \HeII. Finally, we use 32 temporal sub-cycles of the RT solver for each hydrodynamical timestep, and a reduced-speed-of-light approximation (with an effective speed of light $\tilde{c} = 0.01 c$). For the zoom-in runs, we compensate the lack of star formation outside of the high-resolution region by including a contribution from the tabulated UV background of \citet{Faucher-Giguere+2009} in the thermo-chemistry, which concludes reionization by $z\sim6$. Conversely, in the large-box runs, reionization is self-consistently simulated thanks to the RT solver and the source population of forming galaxies.

Unfortunately, the resolution achieved by current galaxy-scale simulations is insufficient to self-consistently simulate all relevant physical processes. For this reason, modern simulations include so-called sub-grid models that strive to mimic the resolution-scale effects of processes occurring on much smaller scales. Among the most important ones are star formation and the detailed ISM structure resulting from the associated feedback processes. The Auriga simulations employ the model of \citet{Springel&Hernquist2003}, where star-forming gas is treated as a mixture of a volume-filling rarefied hot gas surrounding --~in pressure equilibrium~-- cold, dense clouds, where new stars are formed. 
On a practical level, stars are stochastically formed in gas cells above a critical number density of $n_\mathrm{crit} = 0.13 \, \mathrm{cm}^{-3}$, with a gas consumption time-scale  of $\tau_\mathrm{SF} = 2.2 \, \mathrm{Gyr}$ at the threshold density, and a shorter timescale at higher densities following a Schmidt-like star formation law. 
Each stellar particle in the simulations consists of a single stellar population with a \citet{Chabrier03} initial mass function. At each timestep, the mass loss and metal return from SNII (assumed to occur instantaneously for stars above $8 \, \mathrm{M}_\odot$), SNIa and asymptotic giant branch (AGB) stars are distributed over neighbouring cells using a top-hat kernel. Their energetic feedback motivates the inclusion of an empirical model for galactic winds, implemented by launching `wind particles' from the stellar particles in a random direction, with a velocity determined by the local dark matter velocity dispersion. Their momentum and mass content is recoupled to the gas just outside the star-forming phase to realize a non-local momentum feedback that emulates the creation of a wind by the multi-phase ISM.

Finally, we employ the FOF+SUBFIND algorithms \citep{SUBFIND} to perform on-the-fly halo and subhalo identification. We require structures to have at least $32$ bound particles to be considered in the analysis.

\begin{table}
    
        \begin{tabular}{lccl}
        \hline
        Simulation & $\Lbox$   & $N_\mathrm{p}$ & Magnetic seeding\\
        name       & $(\hMpc)$ &                & and notes       \\
        \hline
        
        \aucosmo    & $67.74$ &  L4 & cosmological               \\
        \aubier     & $67.74$ &  L4 & Biermann battery           \\
        \audurr     & $67.74$ &  L4 & Durrive battery            \\
        \audurrlt   & $67.74$ &  L3 & Durrive battery            \\
        \ausne      & $67.74$ &  L4 & SNe inj. ($f_B = 10^{-4}$) \\
        \ausneH     & $67.74$ &  L4 & SNe inj. ($f_B = 10^{-3}$) \\
        \hline
        
        \fbcosmo    & $25$ &  $2 \times 256^3$ & cosmological               \\
        \fbcosmonort& $25$ &  $2 \times 256^3$ & cosmological, no RT        \\
        \fbbier     & $25$ &  $2 \times 256^3$ & Biermann battery           \\
        \fbdurr     & $25$ &  $2 \times 256^3$ & Durrive battery            \\
        \fbsne      & $25$ &  $2 \times 256^3$ & SNe inj. ($f_B = 10^{-4}$) \\
        \fbsneH     & $25$ &  $2 \times 256^3$ & SNe inj. ($f_B = 10^{-3}$) \\
        \hline
        \end{tabular}

    \caption{Details of the simulations employed in this paper. From left to right: name, box size, number of particles, seeding mechanism and other details. The top half refers to zoom-in simulations (dubbed \textsc{Au}), while the bottom half is for the simulations with uniform resolution (dubbed \textsc{FB}). For the former set, the resolution follows the nomenclature of the Aquarius project.
    }
    \label{table:simulations}
\end{table} 

\begin{figure*}
    \centering
    \includegraphics[width=\textwidth]{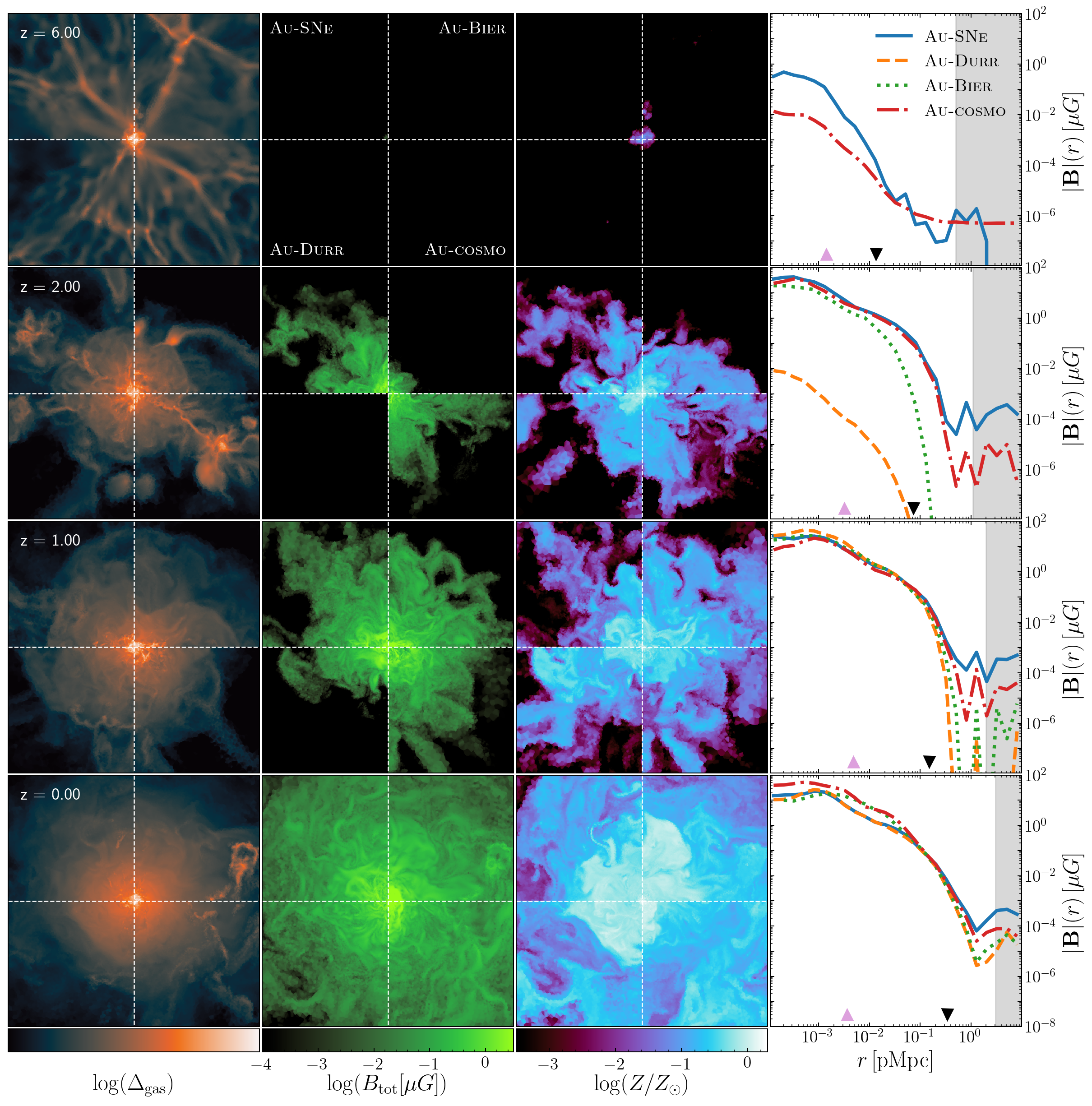}
    \caption{Gas overdensity (in arbitrary units, left column), magnetic field strength (center left) and metallicity (center right) maps in a thin slice centred on the main halo of our \textsc{Au} simulations, with side length of $400 \, h^{-1} \, \mathrm{pkpc}$. Each quadrant of each panel shows results for a different magnetic seed model, as indicated in the top row. Plots in the right column show the \enrico{volume}-averaged magnetic field strength as a function of radius from the centre of the halo for each model. The black lower-pointing (purple upper-pointing) triangle indicates the virial (half-stellar-mass) radius, while the shaded gray region denotes radii outside the highest-resolution region of the zoom-in simulation. Each row shows a different redshift, as reported in the panels of the leftmost column.}
    \label{fig:dens_Bfield_Auriga}
\end{figure*}   

\subsection{Cosmology and initial conditions}

In this paper we assume a \citet{Planck2014} cosmology, \ie $\Omega_\mathrm{m} = 0.307$, $\Omega_\mathrm{b} = 0.048$, $\Omega_\Lambda = 0.693$, and a Hubble constant of $H_0 = 100 \, h \, \kms \, \mathrm{Mpc}^{-1}$ where $h = 0.6777$.
The initial conditions for the zoom-in simulations coincide with halo $6$ of the Auriga project, at their level-$4$ resolution \citep[following the nomenclature introduced in the Aquarius project by][]{Aquarius}, corresponding to a dark matter mass resolution of $m_\mathrm{DM,HR} = 1.97 \times 10^5 \, \hMsol$ in the best-resolved region, which is gradually degraded to $m_\mathrm{DM,LR} = 9.39 \times 10^{10} \, \hMsol$ in the background, and a gas target mass of $m_\mathrm{gas,HR} = 3.83 \times 10^4 \, \hMsol$. We also perform a run with the Durrive seeding model where $m_\mathrm{DM,HR}$ is reduced by a factor of $8$, corresponding to level-$3$ resolution. 
The haloes in the Auriga project were selected to lie in the mass range $1 < M_{200}/10^{12} \mathrm{M}_\odot< 2$, and to be tidally isolated \citep[for more details see][]{Auriga}.

For the \textsc{FB} runs we produce the initial conditions with the \music \citep{MUSIC} code using second-order Lagrangian perturbation theory. We sample the dark matter density field of a cubical, periodic region of the Universe with side length $\Lbox = 25 \, \hMpc$ using $256^3$ particles, corresponding to a mass resolution of $m_\mathrm{DM} = 6.7 \times 10^7\, \hMsol$ and a gas target mass of $m_\mathrm{gas} = 1.3 \times 10^6 \, \hMsol$. All initial conditions are produced at a starting redshift $z_\mathrm{init} = 127$. In Table~\ref{table:simulations}, 
we give a brief summary of the simulations used in this work.

\section{Galactic magnetic fields}
\label{sec:galactic_magnetic_fields}
We start our analysis by investigating the creation and amplification of magnetic fields on galactic scales using the \textsc{Au} simulations. We do so for three main reasons: (i) magnetic fields are observed mainly in galaxies and galaxy clusters, (ii) the dynamo amplification of seed fields primarily occurs in galaxies, and (iii) concentrating numerical resources into a single simulated object allows us to reach much higher resolution than possible with traditional simulations. This is crucial since the turbulent dynamo amplification is progressively faster on smaller scales, and hence its amplification rate is typically resolution-limited in galactic simulations. Therefore the time needed to reach saturation is longer at lower resolution, to the point that a too-coarse resolution may severely under-estimate the magnetic field strength. Note, however, that even state-of-the-art zoom-in simulations still underestimate the amplification rate because of resolution limits.

\subsection{Magnetic field growth}
\label{sec:mag_growth}

An initial overview of our results is presented in Fig.~\ref{fig:dens_Bfield_Auriga}, where we plot thin slices through our \textsc{Au} simulations (showing, from left to right, gas overdensity, magnetic field strength, and metallicity), as well as the radial profile of magnetic field strength (right panels, the shaded bands indicate the region outside the highest-resolution patch). Both the slices and the radial profiles are centred on the main simulated halo. Each quadrant of the slices shows a different magnetic field seeding mechanism (reported in the second top panel from the left), and each row corresponds to a different redshift (annotated in the left panels). This means that quadrants are not pixel-by-pixel comparable, but thanks to the approximate spherical symmetry around the face-on galaxy, the layout provides a compact way to compare the different seeding models.

A few important insights can be gained already from this simple visualisation. For instance, by comparing the quadrants of each panel in the left column, it can be appreciated that the evolution of the simulated galaxy is very similar in all the seeding mechanisms investigated, since no discontinuity across quadrants can be seen. 
\citep[see for instance][for a study of the impact of magnetic fields on galaxy evolution]{Pakmor&Springel2013, Pakmor+2017, vandeVoort+20}.

Unlike the gas density, the magnetic field evolution is very different in the four models investigated. From the top panel it can be seen that \ausne has the largest magnetic field at high redshift. The reason lies in the fact that depositing part of the SN energy into magnetic fields produces seed strengths which are significantly larger at high redshift than in the other seeding models investigated. The latter, additionally, magnetise inter-galactic gas (see Sec.~\ref{sec:gas_density}), that however requires some time to be accreted and amplified within virialised structures, whereas SN deposit the seed field already within galaxies.

This can be quantitatively appreciated in the right column. At $z=6$, the central magnetic field strength is approximately two orders of magnitude larger in the \ausne run with respect to the \aucosmo simulation. Already by $z=2$, the \ausne and \aucosmo runs produce  very similar profiles. However, in the \aubier simulation, although the innermost region of the galaxy has a magnetic field comparable to that from cosmological and SN-injected seeds, $|\mathbf{B}|$ shows a much sharper drop with radius (appreciable also in the second column), as a consequence of the delayed onset of the Biermann battery. In fact, since magnetic field amplification is most efficient at early times in the very centre of the galaxy \citep{Pakmor+2017}, saturation will be reached first in such a region, and only at later time at larger radii (either because of in-situ amplification or migration of highly-magnetised gas from the galaxy centre). 
As time progresses, the magnetic field strength settles to very similar values in all runs, as a consequence of amplification mechanisms taking over and erasing any memory of the original seed. While at $z=1$ there are still minor differences at the largest scales ($r \gtrsim 1 \, \mathrm{pMpc}$), by $z=0$ these are also completely erased.
In the very central regions of the galaxy, the magnetic strength is larger for the \aubier and \aucosmo runs, although only by a factor of approximately $2$. In light of the results presented in Sec.~\ref{sec:haloes}, we attribute this difference to the stochasticity in the galaxy formation process, in this case triggered by the different times at which the magnetic field becomes relevant in runs with different seeding (see also Fig.~\ref{fig:B_center_evol_Auriga}).

\begin{figure}
    \centering
    \includegraphics[width=\columnwidth]{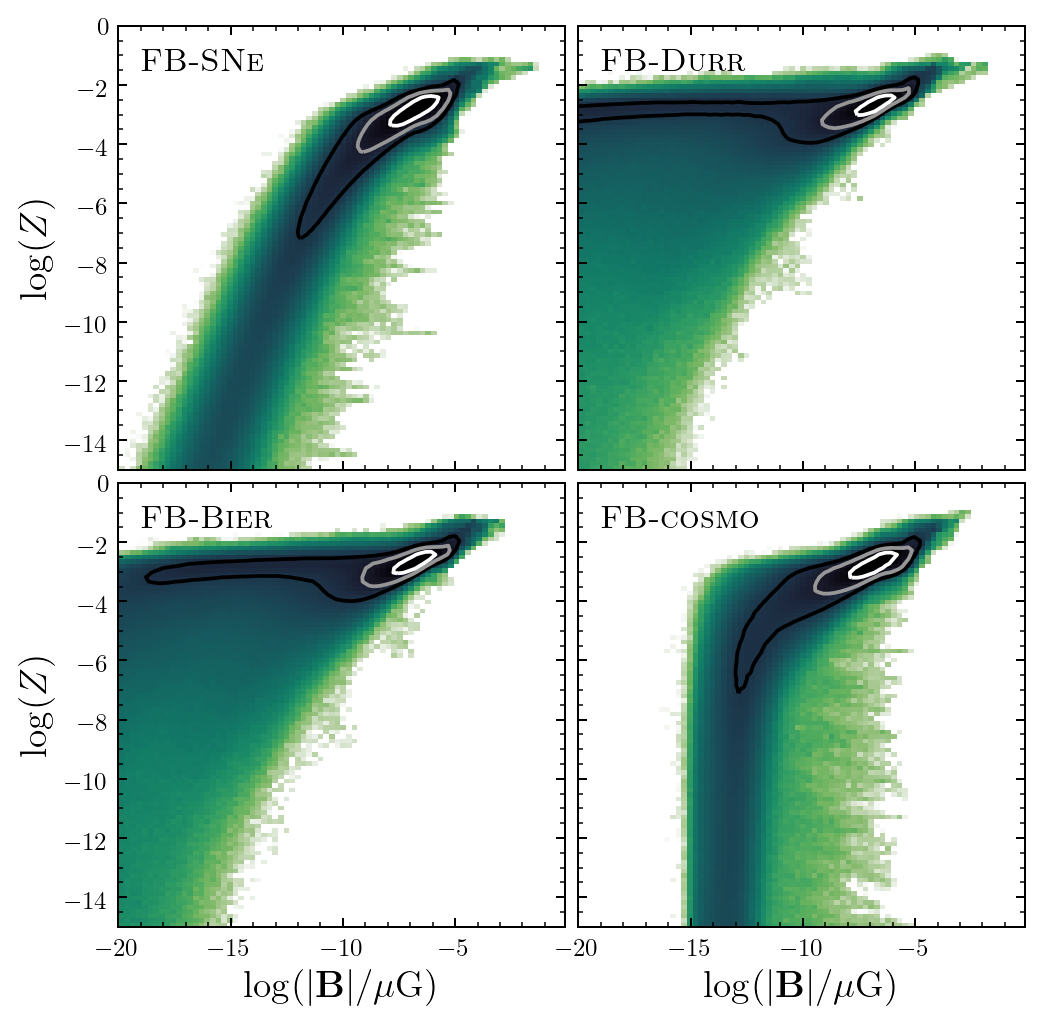}
    \caption{Two-dimensional histogram of magnetic field strengths and gas metallicity in all the simulated gas cells in the \textsc{FB} simulation set at $z=0$. Each bin is color-coded (from light to dark) to reflect the total gas mass that falls in it. The contours indicate 50\%, 70\%, and 90\% of the maximum value.}
    \label{fig:B_vs_Z_z0}
\end{figure}

\begin{figure}
    \centering
    \includegraphics[width=\columnwidth]{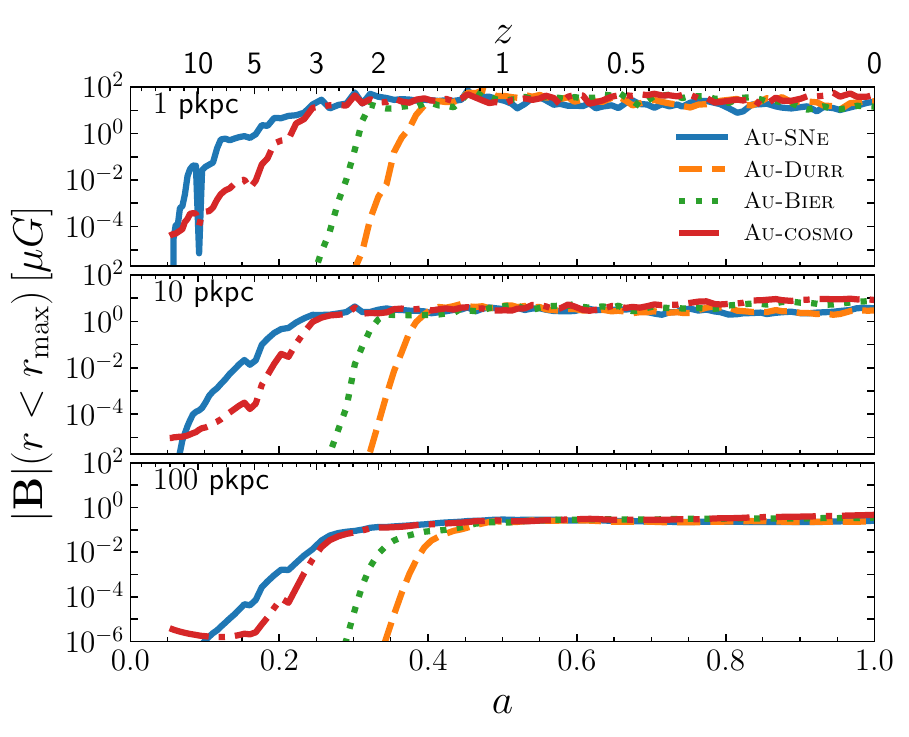}
    \caption{Evolution of the volume-weighted magnetic field strength within a spherical region of three different radii, namely (from top to bottom) $1$, $10$, and $100 \, \mathrm{pkpc}$, centred around the main high-resolution halo in the \textsc{Au} simulation set.}
    \label{fig:B_center_evol_Auriga}
\end{figure}

\begin{figure}
    \centering
    \includegraphics[width=\columnwidth]{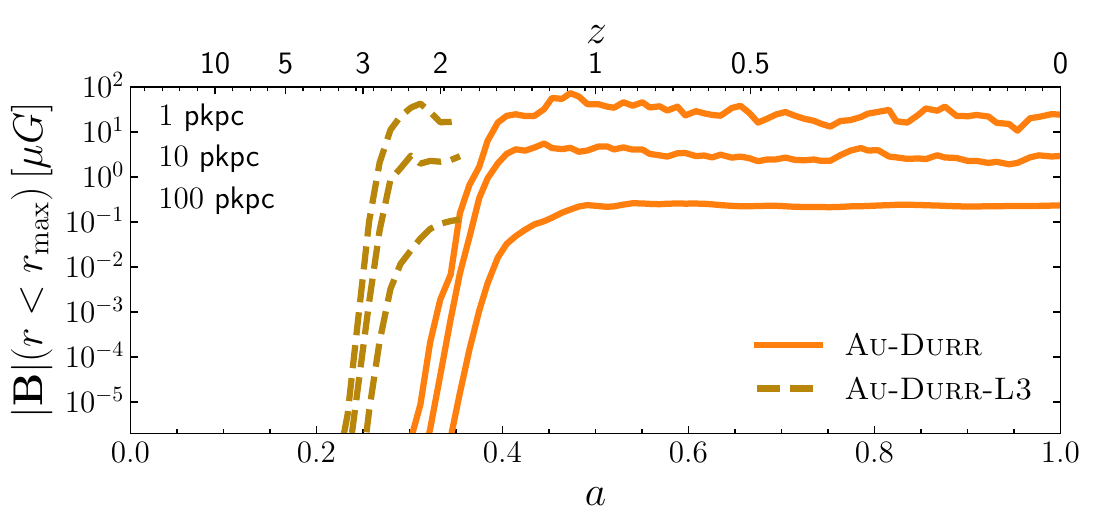}
    \caption{Similar to Fig.~\ref{fig:B_center_evol_Auriga}, but comparing the fiducial- (solid lines) and higher-resolution (dashed lines) runs with the Durrive battery as the magnetic seeding scheme. Each pair of solid -- dashed lines corresponds to a different sphere radius, as reported in the top left corner of the figure in the same order as the curves.}
    \label{fig:res_B_center_evol}
\end{figure}

Comparing the distribution of metals (third column from the left) with the one of magnetised gas, it can be appreciated that, at late times, there is a spatial correlation between magnetic field strength and metallicity. 
\enrico{This is shown more quantitatively in Fig.~\ref{fig:B_vs_Z_z0}, where we report the two-dimensional distribution of magnetic field strength and metallicity in the gas simulated as part of the \textsc{FB} set (we obtain similar results from the high-resolution region of the \textsc{Au} simulations), with a color-coding reflecting (from light to dark) the total gas mass in each bin. Contours enclose regions above 50\%, 70\%, and 90\% of the maximum value, respectively. The correlation discussed above is clearly visible in the top right part of each histogram, and is in fact very similar in all models investigated. As discussed above, this is not the case at earlier time, when the original position and strength of the magnetic seeds play an important role. In particular, the correlation is already in place at $z\sim6$ in the \fbsne and \fbcosmo runs, while it is established only at much later time ($z \lesssim 2$) for the other two \textsc{FB} runs. The sharp drop in the bottom-right panel of the Figure for $\log(|\mathbf{B}| / \mu \mathrm{G}) \lesssim -15$ is simply due to the uniform seed field (combined with the effect of flux conservation during the expansion of the Universe), setting a universal minimum value for the magnetic field.}

\enrico{The origin of this spatial co-presence has been thoroughly investigated in \citet{Pakmor+2020} employing dedicated simulations with a setup very similar to the one used in this work, but forcing high numerical resolution in the CGM, in order to better investigate the evolution of magnetic field \textit{around} galaxies. The authors showed that the correlation discussed here is the result of the co-spatial metal enrichment and magnetic amplification occurring within galaxies. Feedback processes \textit{subsequently} drive the metal-enriched and highly-magnetised gas to larger distances. In haloes with total mass $M_\mathrm{halo} \lesssim 10^{12} \, \hMsol$, the almost complete absence of amplification and enrichment processes outside of the galaxies entails that the spatial correlation between high metallicity and strong field strengths is maintained for a long time. The situation may be different in larger structures \citep[see \eg][although employing non-radiative simulations, which lack a realistic galaxy population and hence do not allow us to estimate the relative importance of \textit{in situ} amplification and feedback-driven ejection of magnetic fields]{Vazza+18}.
However, since the magnetic field amplification within galaxies critically depends on the initial strength and injection time of the seed field, which vary wildly in the models tested, the time when this correlation is established can vary significantly (\ie for very weak seed fields, at earlier time the feedback processes expel metal-enriched gas that is not yet highly magnetised). }

In order to better visualise \enrico{the different time of galactic and circum-galactic magnetisation}, we show in Fig.~\ref{fig:B_center_evol_Auriga} the volume-averaged magnetic field strength in spherical regions with different radii (from top to bottom, $1$, $10$, and $100 \, \mathrm{pkpc}$) centred on the main galaxy in the \textsc{Au} simulations. Different mechanisms reach saturation at different times, with \aubier and \audurr being the last to do so. This opens up the possibility of distinguishing the dominant seeding mechanism using observations of magnetic fields in high-redshift galaxies. While difficult in practice, there are few examples of such measurements in the literature. Most notably, \citet{Kronberg+2008} employed a set of $900$ measurements of extra-galactic Faraday rotation extending to $z\approx 3.7$ in order to determine that the observed distribution of RMs can only be explained by (i) the presence of magnetic fields with present-day strength in galaxies above $z\gtrsim1.5$, or (ii) a large number of magnetised clouds in the inter-galactic medium. Both putative explanations are somewhat hard to reconcile with the Durrive injection model, since in our simulations the magnetic fields it generates reach their saturation level at approximately $z\approx 1.5$ in galaxies. 
In a subsequent work, \citet{Bernet+2008} showed that the largest RM are statistically associated with the largest MgII absorption. Since the latter occurs in and around galaxies, its association with large RM indicates that explanation (i) above is likely the correct one.

However, it should be noted that the amplification rate increases with resolution \citep[see \eg figure 16 of][]{Pakmor+2017}. Hence, the saturation time in our simulations should be considered an upper limit. To confirm that this is the case in our simulations as well, we have re-run the \audurr simulation with resolution improved by a factor of $8$ (dubbed \audurrlt in Table~\ref{table:simulations}). In Fig.~\ref{fig:res_B_center_evol}, we compare the volume-averaged magnetic field strength evolution in the two runs. With better resolution, the amplification mechanisms are indeed more efficient, and the magnetic field strength reaches its saturated value earlier in the \audurrlt run (dashed lines) than in the lower resolution \audurr simulation (solid lines). 

\enrico{
At $z=0$, in all models investigated the ratio between the magnetic ($\varepsilon_\mathrm{mag}$) and turbulent\footnote{Note that we remove the large-scale ordered gas motion from the calculation of this quantity by computing $\varepsilon_\mathrm{turb}$ as three times the value obtained from the gas motion perpendicular to the galactic disk only. While approximate, this method in our experience gives relatively accurate results. More importantly, as we are here focusing on the inner region of the galaxy, turbulent gas motion is dominant over the ordered one.} ($\varepsilon_\mathrm{turb}$) energy densities in the 
central $2$~kpc (physical) of the simulated galaxy is approximately $\varepsilon_\mathrm{mag} / \varepsilon_\mathrm{turb} \approx 0.22$ (with model-to-model variation of order $0.02$), consistent with previous numerical studies of the Auriga project \citep{Pakmor+2017}. This value also corresponds to the one expected from MHD turbulence simulations of the ISM once the turbulent dynamo saturates \citep[see \eg][]{Cho+2009,Kim&Ostriker15, Federrath16}, and is compatible with measurements in the Milky Way centre from \citet{Crocker+10}. 
}

\subsection{Impact on galaxy properties}

\begin{figure}
    \centering
    \includegraphics[width=\columnwidth]{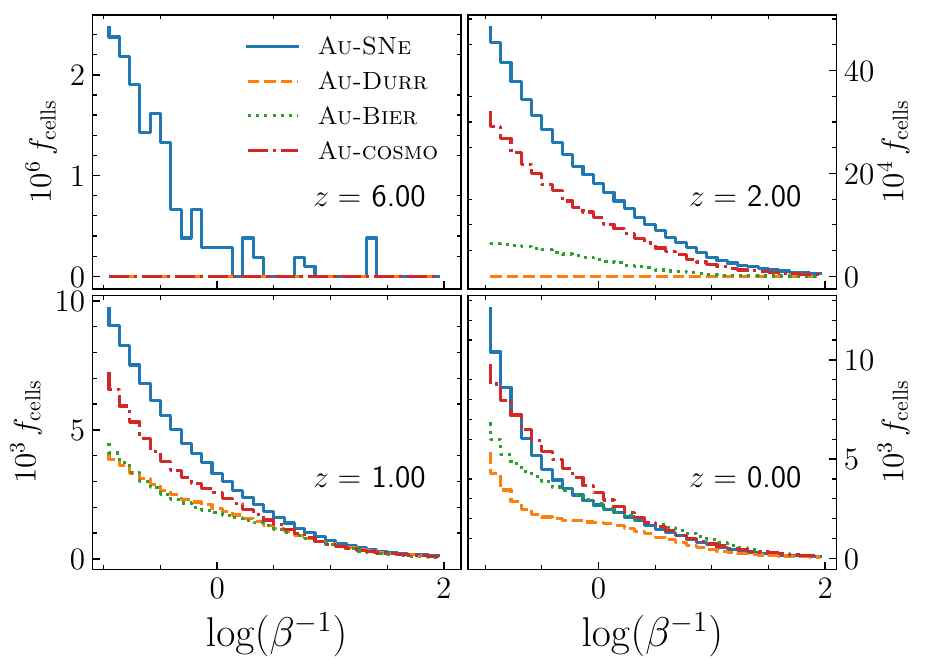}
    \caption{Distribution of magnetic-to-thermal pressure ratio $\beta^{-1} \equiv P_\mathrm{magnetic} / P_\mathrm{thermal}$ in the high-resolution region of the \textsc{Au} simulations, at $z=0$, $1$ , $2$, and $6$. All the histograms are normalised to the total number of gas cells in the simulation (note the different vertical normalizations in the plot).}
    \label{fig:histogram_beta_inv_Auriga}
\end{figure}

The distinct amplification times in different seeding mechanisms imply that the relevance of their magnetic pressure $P_\mathrm{magnetic} \equiv (8 \pi)^{-1} \, |\mathbf{B}|^2$ (in CGS units) for galaxy formation may be different. We explore this possibility in Fig.~\ref{fig:histogram_beta_inv_Auriga}, where we show the fraction of cells in the high-resolution region of the \textsc{Au} simulation as a function of their magnetic-to-thermal pressure ratio $\beta^{-1} \equiv P_\mathrm{magnetic} / P_\mathrm{thermal}$ at $z=6$, $2$, $1$, and $0$ (note that, for visualization purposes, we employ a different normalization in each panel). The \ausne simulation shows the largest number of cells with large $\beta^{-1}$ at almost every redshift investigated, although the fraction of gas cells with $\beta^{-1} > 1$ is very small ($\sim 10^{-6}$ -- $10^{-3}$), suggesting a limited impact on the evolution of the galaxy itself. 

As a simple test of the contribution of magnetic fields to galaxy evolution, we report in Table~\ref{table:gal_prop}, for the main galaxy in all the \textsc{Au} simulations (including one with no seeding mechanism enabled, \ie \aunone), some basic galaxy properties at $z=0$. We estimate the halo virial 
mass (listed in the second 
column from the left
) from the spherical region --~centred on the galaxy~-- with density equal to $200$ times the mean density of the Universe. The stellar half-mass radius and the stellar mass it encloses\footnote{We note that in the original Auriga paper \citep{Auriga} the galaxy stellar mass was defined as the one contained within $0.1$ times the virial radius. We have used a different measure to better capture the effect of magnetic fields on the stellar distribution, since we do not define the latter based on halo properties, which are mostly driven by the dark matter distribution.} are shown in the fourth and fifth column of the Table. Finally, we report in the 
two 
rightmost 
columns the volume-averaged magnetic field strength in the central $10 \, \mathrm{kpc}$ of the galaxy and within the galactic disk (defined as the cylinder with radius equal to twice the stellar half-mass radius and a height of $2\,\mathrm{kpc}$ centred on the galaxy centre).

The galactic properties detailed in Table~\ref{table:gal_prop} appear to be very similar in all the magnetic seeding model investigated, with few exceptions. The stellar half-mass radius is slightly larger in the \ausne model, while the enclosed mass is in line with the other \textsc{Au} runs, indicating a moderately less-concentrated stellar component. This is consistent  with the greater (although limited) impact of magnetic fields on gas reported in Fig.~\ref{fig:histogram_beta_inv_Auriga}. The other significant difference between the models investigated is the strength of the central and disk magnetic field, which is higher in the \aubier and \aucosmo runs by approximately a factor of $2$ -- $3$. 
As already discussed, we believe the difference is due to stochasticity.
Overall, however, the effect of all magnetic seeding mechanisms on the properties of the galaxies appears mild. Even the differences in magnetic field strength in the central region pales compared to the uncertainties in its measured values.

We have shown in this Section that individual simulated galaxies appear very similar at $z\lesssim1$, irrespectively of how magnetic fields are generated. Although we have only investigated one object, it has been shown in \citet{Pakmor+2016} that galaxies in the Auriga suite do not exhibit extreme variability in their magnetic properties. We are therefore confident that conclusions drawn from the investigation presented here are general, at least within the Auriga simulation suite. We confirm this next by studying the magnetic fields in galactic haloes from the \textsc{FB} simulations.

\begin{table}
    \begin{tabular}{lccccccc}
    \hline
    Simulation & 
    
    $\Mtm$ & 
    $\Rhms$ 
    & $\Mhms$ 
    & $|\mathbf{B}_{<10\,\mathrm{kpc}}|$ 
    & $\mathbf{B}_\mathrm{disk}$\\
    
    name & 
    
    $[10^{10}\,\mathrm{M}_\odot]$ & 
    $[\mathrm{kpc}]$ & 
    $[10^{10}\,\mathrm{M}_\odot]$ & 
    $[\mu\mathrm{G}]$ & 
    $[\mu\mathrm{G}]$\\
    
    \hline   
    
    \aunone  
              & $129.60$ & $4.29$ & $1.31$ & $0.00$ & $ 0.00$ \\ 
    \ausne   
              & $133.05$ & $5.21$ & $1.45$ & $2.94$ & $ 3.35$ \\ 
    \ausneH  
              & $133.29$ & $6.04$ & $1.47$ & $3.62$ & $ 3.49$ \\ 
    \audurr  
              & $131.76$ & $4.79$ & $1.43$ & $2.95$ & $ 3.61$ \\  
    \aubier  
              & $133.22$ & $4.49$ & $1.63$ & $8.28$ & $ 8.12$ \\ 
    \aucosmo 
              & $133.91$ & $3.64$ & $1.65$ & $8.30$ & $15.57$ \\ 
    \hline
    \end{tabular}
    
    \caption{Some halo and galaxy properties in the \textsc{Au} simulations at $z=0$. From left to right: halo virial 
    mass, galaxy stellar half-mass radius, mass enclosed in the latter,  volume-averaged magnetic field strength within $10 \, \mathrm{kpc}$ from the galaxy centre and within the galactic disk (defined as the cylinder with radius equal to twice the stellar half-mass radius and a height of $2\,\mathrm{kpc}$ centred on the galaxy centre).}
    \label{table:gal_prop}
\end{table}

\subsection{Magnetic fields in haloes}
\label{sec:haloes}

\begin{figure}
    \centering
    \includegraphics[width=\columnwidth]{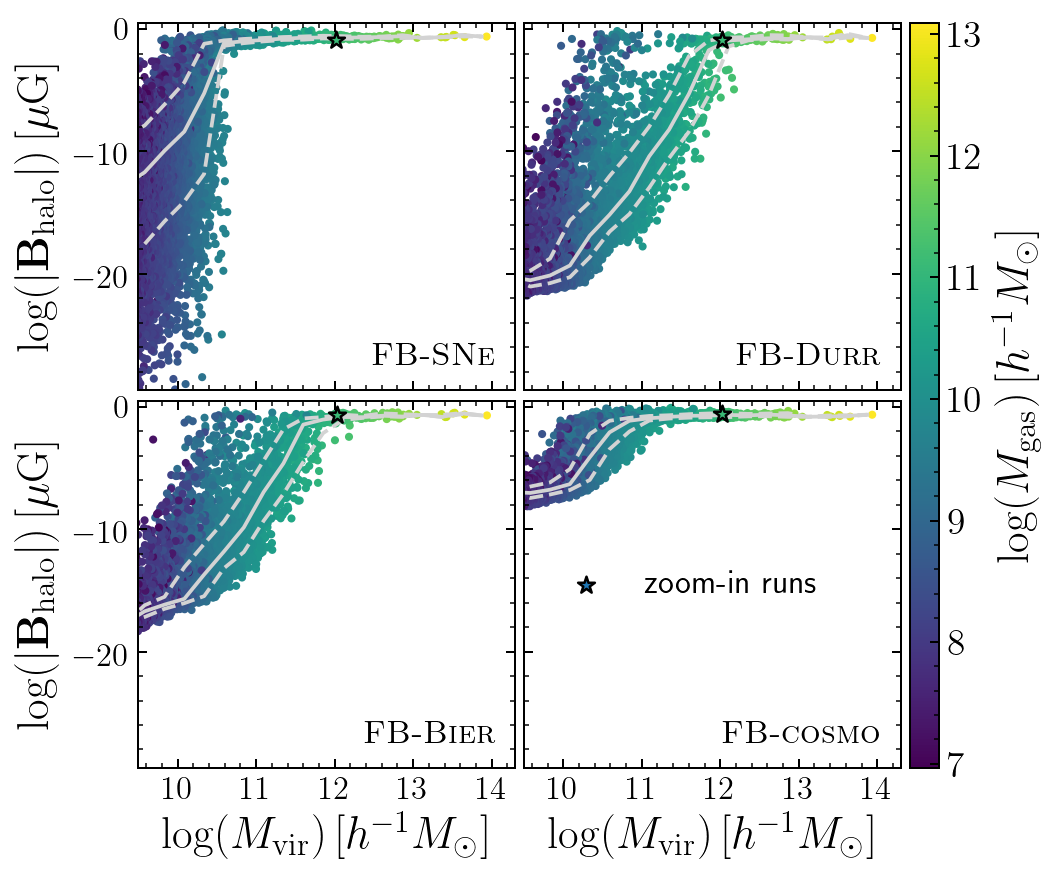}
    \caption{Volume-averaged halo magnetic field strength as a function of halo virial mass. Each point is colour-coded following the halo gas mass. The grey solid and dashed lines report the median and $16$th/$84$th percentiles of the distribution. Finally, the star symbols show the \textsc{Au} runs with the corresponding seeding mechanism, using the same colour scheme.}
    \label{fig:B_vs_mvir_z0}
\end{figure}

In order to study the magnetic fields in the simulated galaxy populations, we compute the volume-averaged magnetic field strength within the virial radius of all haloes in the \textsc{FB} runs. In Fig.~\ref{fig:B_vs_mvir_z0} we show this quantity 
as a function of halo virial mass. Each circle corresponds to a simulated halo, and its colour reflects the halo gas mass. In each panel, the corresponding \textsc{Au} run is also shown using a star symbol (and the same colour-scheme).  Before moving on, we note here that, while the amplification rate is resolution-limited in our simulations, the relative differences among models are expected to be robust. 

The distributions in the four panels appear to have some common features. Firstly, in all runs the magnetic field strength saturates at the same value of approximately $1 \mu\mathrm{G}$ in massive haloes, indicating once again that this is primarily determined by the structure formation process \citep[see also \eg][]{Marinacci&Vogelsberger2016}\enrico{ rather than the initial seed field. It should be noted that, for the purpose of this discussion (which mainly focuses on magnetic seeding processes), we consider the feedback from small-scale structures (\eg supernov\ae) to be part of the structure formation process, as it is an unavoidable ingredient of any successful simulation including baryonic matter. \sout{The latter} Structure formation} is also \enrico{partly }responsible for the existence of a seeding-dependent virial mass threshold for saturation. In fact, larger haloes are more efficient in the amplification process since: (i) they reach larger gas overdensities, and hence stronger magnetic fields through flux conservation, (ii) they undergo more merger events, which excite shear flows and turbulence, \ie two of the primary mechanisms of field amplification, (iii) additional physical processes \enrico{\sout{like ram-pressure stripping}} can generate additional turbulence in large haloes \citep{Marinacci+2015}\enrico{, (iv) the Reynolds number scales with numerical resolution $\Delta x$ and halo virial radius $R_\mathrm{200}$ as $\mathrm{Re} \approx (R_\mathrm{200}/\Delta x)^{4/3}$, and hence is larger in larger haloes, rendering the amplification of magnetic field faster}. Hence, for any given amount of time and fixed magnetic seed model, saturation will be reached first in larger structures. 

The threshold halo mass for saturation appears to be different in the models investigated. More specifically, in the \fbsne run all haloes with virial mass above approximately $10^{10.7}\,\hMsol$ have saturated magnetic field, while this threshold moves to $\sim10^{11}\,\hMsol$ in \fbcosmo and $10^{12}\,\hMsol$ in \fbbier and \fbdurr, respectively. The reason for this difference is a combination of two different effects: (i) a lower initial seed field requires longer amplification times until saturation, hence necessitating a larger halo mass to reach saturation in any given amount of time, and (ii) different magnetic seeding prescriptions act at different times, granting haloes an unequal amount of time to amplify the magnetic field strength (see also Fig.~\ref{fig:vol_avg} and its discussion).

In the low-halo-mass end of Fig.~\ref{fig:B_vs_mvir_z0}, the \fbsne run shows a considerably larger scatter than the other runs (even when normalised for the median magnetic field strength, not shown in the Figure). The reason lies in the very broad distribution of magnetic field strengths before entering the haloes (see Sec.~\ref{sec:gas_density} for a discussion), as well as in the stochasticity of star formation --~and, hence, of SN events~-- at low masses. In fact, at the resolution of our simulations, haloes below approximately $M_\mathrm{vir} \sim 10^{11} \, \hMsol$ form only a handful of stellar particles. While the latter is just a consequence of resolution, it may indicate that galaxies hosting only a handful of stars retain traces of the magnetisation of the gas they originate from. This could have interesting implications for constraining inter-galactic magnetic fields  (and hence seeding models) using ultra-diffuse galaxies. However, tailored simulations are needed to investigate this hypothesis in more detail.

Since haloes are the primary locations of magnetic field amplification, it can be expected that the different timing and amount of saturated gas produce differences in the large-scales magnetic fields. In order to move forward in this direction, we study in the next section the magnetic properties of the Universe as a whole.

\section{Large-scales magnetic fields}
\label{sec:large_scales_magnetic_fields}

\begin{figure*}
    \centering
    \includegraphics[width=\textwidth]{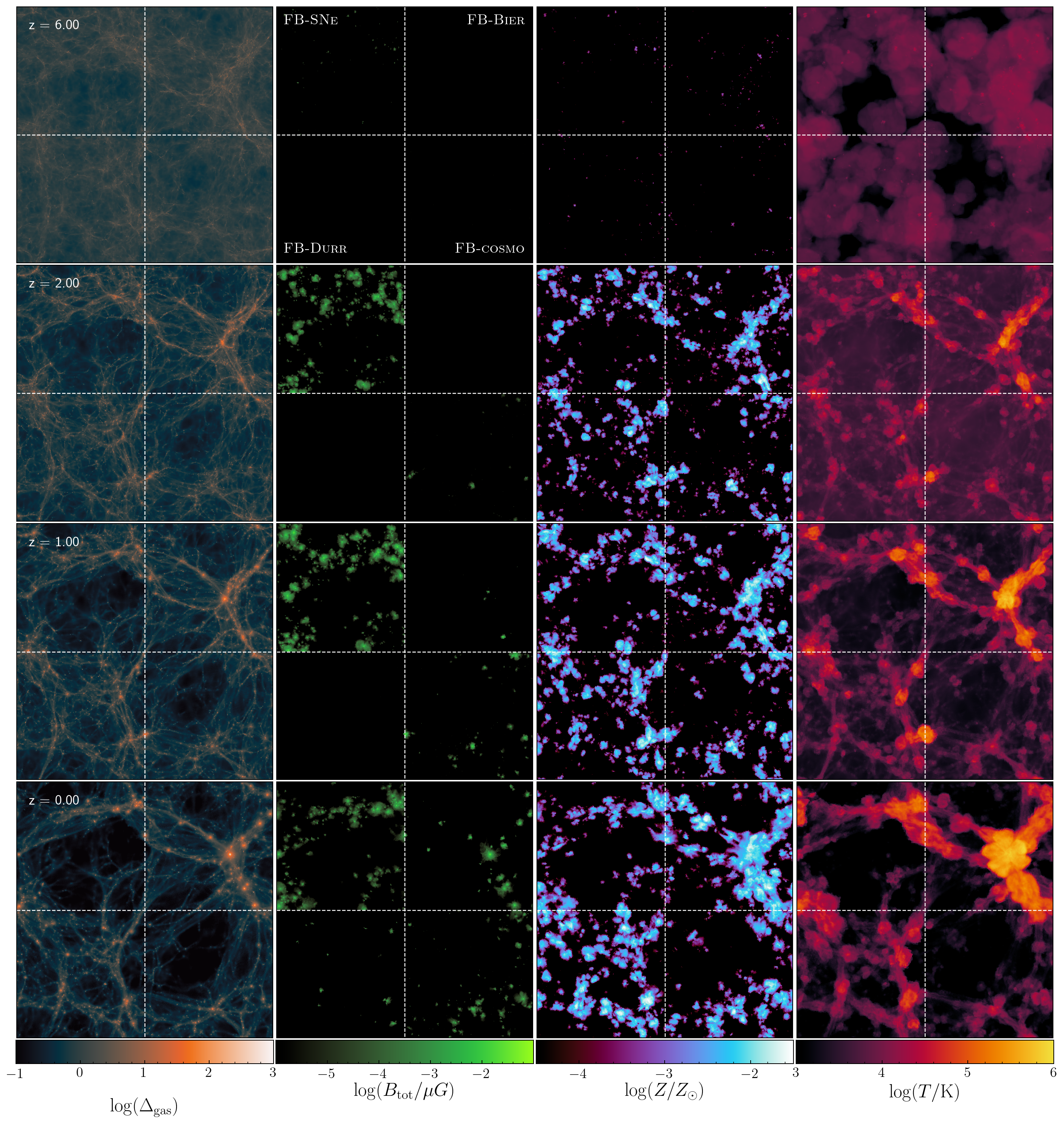}
    \caption{Similar to Fig.~\ref{fig:dens_Bfield_Auriga}, but for the \textsc{FB} runs. Here, the rightmost column shows the gas temperature, and the slices span the entire simulation box in the two dimension shown, and half of the box is projected along the viewing direction.}
    \label{fig:split_panels_fb}
\end{figure*}

In the previous section we have shown that all the different magnetic seeding mechanisms investigated in this paper are able to generate virtually indistinguishable magnetic fields at $z=0$ in and around Milky Way-like galaxies, as well as similar degrees of halo gas magnetisation (once saturation is reached). However, the location where magnetic seed fields are created in the gas differs significantly, potentially leaving large-scales signatures behind. 
In Fig.~\ref{fig:split_panels_fb}, we provide an overview of the simulated gas in a similar fashion to Fig.~\ref{fig:dens_Bfield_Auriga}. Each column shows a different gas quantity, namely from left to right: density contrast (defined as $\Delta_\mathrm{b} \equiv \rho_\mathrm{b} / \langle \rho_\mathrm{b} \rangle$, where $\rho_\mathrm{b}$ is the density of baryons), magnetic field strength, metallicity \citep[in solar units, where the solar metallicity is $Z_\odot = 0.0127$, ][]{Wiersma+09}, and temperature. 
All panels report mass-weighted projections of a slab that extends for the entire simulation box in two dimensions, and for half of it in the third (projected) dimension. 

From Fig.~\ref{fig:split_panels_fb} it can be seen that the gas density, metallicity, and temperature 
are unaffected by the seeding mechanism, indicating that all of them provide a similar  (potentially negligible) degree of magnetic pressure support. 
In fact, 
the magnetic field outside of galaxies remains always very small. The latter is discussed more quantitatively in the next section, but can also be appreciated in the second column from the left. The \fbsne run produces relatively strong (of order $0.1 \, \mu \mathrm{G}$) magnetic fields earlier than any other model. As in the \textsc{Au} runs, the reason is twofold: (i) this seeding mechanism injects the locally strongest magnetic fields, and (ii) seed fields are injected directly in galaxies. Both of these features aid the amplification of the magnetic field. Early magnetisation leaves more time for the magnetised gas to be expelled from galaxies and to reach the surrounding environment. This results in more extended regions of appreciable magnetic field strength.  
In the next sections we will substantiate all the qualitative observations made above, starting from the magnetic field strength in different cosmic environments.

\subsection{Gas magnetisation -- density relation}
\label{sec:gas_density}
\label{sec:vol_avg}

\begin{figure}
    \centering
    \includegraphics[width=\columnwidth]{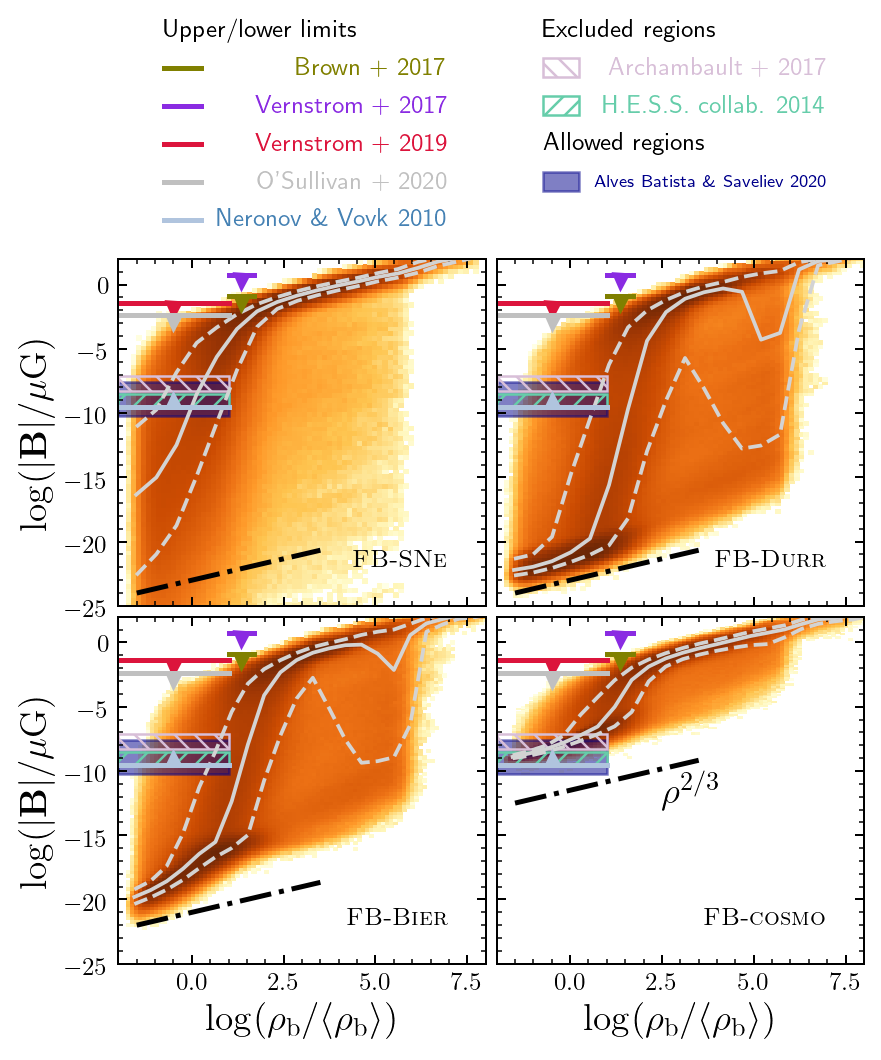}
    \caption{Two-dimensional histogram of magnetic field strengths and gas density in all the simulated gas cells. Each bin is color-coded to reflect the total gas mass that falls in it. The grey solid and dashed lines report the median and $16$th/$84$th percentiles of the distribution. The dot-dashed line shows the expected scaling from pure magnetic flux conservation (with arbitrary normalisation). Arrows shows upper and lower limit on magnetic fields in the IGM and filaments of the cosmic web \enrico{inferred by \citet{Neronov&Vovk2010}, \citet{Vernstrom+17,Vernstrom+19}, \citet{Brown+17}, and \citet{OSullivan+20}, while the hatched patches show the excluded regions from \citet{HESScollab14}, and \citet{VERITAScollab17}. Finally, the shaded region shows the \textit{allowed} magnetic strength from \citet{AlvesBatista&Saveliev20}.} For all these observational studies, we have assumed that the IGM extends to overdensities $\rho / \langle \rho_\mathrm{b}\rangle \lesssim 10$\enrico{, while the cosmic web includes overdensities $10 \lesssim \rho / \langle \rho_\mathrm{b}\rangle \lesssim 50$ \citep[see \eg][]{Tanimura+20}}.}
    \label{fig:B_vs_rho_z0}
\end{figure}

We take 
full advantage of the broad range of cosmic environments simulated in the \textsc{FB} set in Fig.~\ref{fig:B_vs_rho_z0}, where we show at $z=0$ the magnetic field strength as a function of baryonic density (normalised by its average value) as a two-dimensional histogram. The colour intensity (from light to dark) reflects the amount of mass that falls in each bin. Lines indicate the median (solid) and central $64$ percentiles (dashed). Finally, the dot-dashed lines show the scaling expected from pure flux conservation. 

We start by discussing the low and intermediate density regime. In the \fbsne run (top left panel), the initial seed is injected, by construction, at the location of SN events, which always occur inside galaxies. Hence, magnetised gas found in low-density environments at $z=0$ has been previously processed in a galaxy and expelled from it\footnote{Note, however, that not all the gas in low-density environments has been previously in a galaxy, and hence some of it is not magnetised (not shown in Fig.~\ref{fig:B_vs_rho_z0} but discussed in Fig.~\ref{fig:f_magnetised}).}. Therefore, 
its magnetic field varies by large amounts depending on the degree of amplification it underwent before being transported into low-density regions. This is the reason for the large range covered by the 16th-8th percentiles of the distribution. The opposite configuration occurs in the \fbcosmo run (bottom right panel), where the gas starts with a uniform (weak) magnetic field in the initial conditions. In this case, since most of the low-density gas never entered into a halo, conditions are ideal for the flux-conservation scaling to hold. In fact, the scatter in this relation is vanishing at the lowest-density end. An intermediate situation is found in the simulation run including the Durrive battery (top right panel). Most of the low-overdensity gas follows the flux-conservation scaling, but the scatter is larger than in the cosmological seed case. 

Similarly to the \fbsne run, the scatter is produced by a non-uniform initial gas magnetisation, since in this seeding model the latter depends on the properties of matter overdensities and radiation fields around galaxies at cosmic dawn. It is however interesting to note that \textit{there is} a $|\mathbf{B}|$ - $\rho_\mathrm{b}$ relation in low-density gas. This is a consequence of the seeding mechanism being able to produce magnetic fields in low-density gas far away from galaxies, which is less affected by structure formation processes but then appears to depart from pure flux conservation. 
Finally, when magnetic fields are produced via the Biermann battery, the slope of the $|\mathbf{B}|$ - $\rho_\mathrm{b}$ relation in low-density gas is steeper than $2/3$. This is likely a consequence of the battery process itself, which occurs preferentially in filaments during the EoR \citep{Gnedin+2000} and in shocks \citep{Ryu+98}, where additional processes on top of flux conservation are present.

In the Figure we report as a light blue band the lower limit on the IGM magnetic fields from \citet[][for a magnetic field coherence length of order $\lambda_B \sim 1$ Mpc; the value scales as $\lambda_B^{-1/2}$]{Neronov&Vovk2010}. This lower bound was inferred from the missing extended gamma-ray emission (due to particle cascades) around the primary point source. If taken at face value, they are in tension with the results from the \fbdurr and \fbbier runs, since only a small fraction of the simulated IGM has magnetic fields larger than the lower limits (see also Fig.~\ref{fig:f_magnetised} and its discussion). However, \citet{Broderick+12} and subsequent works have shown that such lower limits may not be reliable, since they are based on the assumption that inverse Compton scattering is the primary energy-loss mechanism for the ultra-relativistic particle pairs produced during the particles cascade, whereas plasma instabilities can contribute or even dominate the energy dissipation budget \enrico{\citep[see however \eg][]{AlvesBatista+19}}. 
\enrico{We additionally plot the upper limits from studies of the Faraday Rotation difference between sources at small projected separations \citep{Vernstrom+19,OSullivan+20}, and those derived from the search for synchrotron emission from the cosmic web by \citet{Brown+17} and \citet{Vernstrom+17}.}

Finally, we report as hatched regions the values of intergalactic magnetic field excluded by \citet{HESScollab14} and \citet{VERITAScollab17} from the analysis of the magnetic broadening of the $\gamma$-ray emission from blazars.
These excluded regions are in tension with the results from the \fbsne and \fbcosmo runs. In the former, however, the excluded strengths are reached only at gas overdensity between $1 \lesssim \rho_\mathrm{b} / \langle \rho_\mathrm{b} \rangle \lesssim 10$, where the \citet{HESScollab14} and \citet{VERITAScollab17} may not fully apply, as they are derived for the IGM gas only. More serious is the tension in the \fbcosmo run. In this case, almost all the simulated gas in underdense regions shows magnetic field strengths excluded by available data. As already mentioned \citep[and thoroughly discussed in ][]{Marinacci&Vogelsberger2016}, the magnetic field in the IGM is simply set by the initial seed strength combined with flux conservation. Hence, this tension may be simply resolved by changing the initial seed strength within the allowed parameter space. In this work we employed the same seed strength as in the Auriga and Illustris-TNG simulation suites. However, as abundantly discussed in Sec.~\ref{sec:galactic_magnetic_fields}, the magnetic fields in overdense regions are not affected by this choice.
\enrico{By combining measurements of the time delay between neutrinos and gamma rays from the blazar TXS 0506+056 with the time dependence of its gamma-ray spectrum, \citet{AlvesBatista&Saveliev20} constrained the intergalactic magnetic fields to the values shown with a shaded box in the Figure. Curiously enough, these allowed values overlap with the magnetic strengths excluded by the measurements discussed above, highlighting the need to interpret these observational results with caution.} 

In all models, there exists a critical gas overdensity of $\Delta_\mathrm{b} \approx 100$, above which the amplification mechanisms become dominant and boost the magnetic field by several orders of magnitude until saturation. Above this threshold, the magnetic field strength is very similar in all four models investigated, as a consequence of (saturated) amplification in galaxies erasing memory of the original magnetic seed from the gas. Interestingly there exists a $|\mathbf{B}|$ - $\rho_\mathrm{b}$ relation also in this gas, with a slope very close to the pure-flux-conservation prediction. The reason is not to be found in larger haloes producing stronger magnetic fields (see Sec.~\ref{sec:haloes}). Rather, after the gas has been magnetised until saturation, compression and expansion processes due to gas dynamics change the magnetic field strength through flux conservation, hence establishing an offset in the $|\mathbf{B}|$ - $\rho_\mathrm{b}$ relation. In fact, gas that is highly magnetised at $z=0$ first rises from the lower relation to the upper one in a short time (during its amplification phase), and then moves along the latter until $z=0$. 

\begin{figure}
    \centering
    \includegraphics[width=\columnwidth]{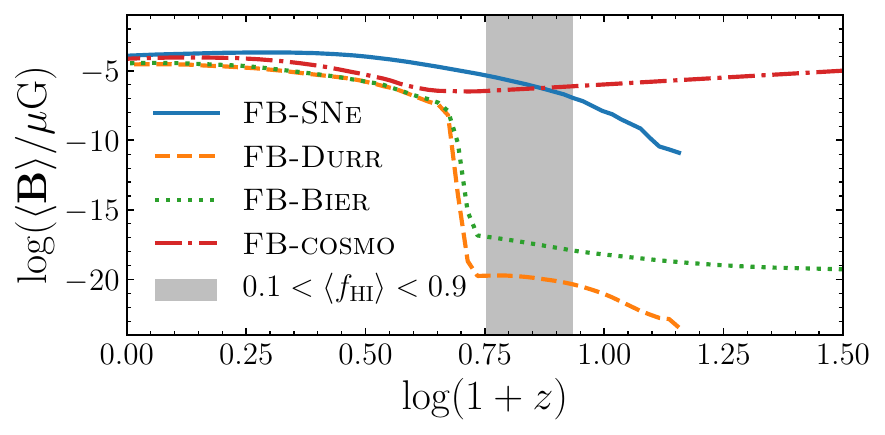}
    \caption{Volume-averaged magnetic field strength in the \textsc{FB} simulations as a function of redshift. The grey vertical band indicates where the hydrogen neutral fraction traverses the interval $[0.1, 0.9]$, roughly bracketing the epoch when most of the ionization fronts are sweeping through the simulated universe. Note that the curves corresponding to \fbsne and \fbdurr start from $z\approx14$, which is the time of formation of the first stellar particles in the simulations.}
    \label{fig:vol_avg}
\end{figure}

In Fig.~\ref{fig:vol_avg} we complement the discussion above by showing 
the volume-averaged magnetic field strength $\langle |\mathbf{B}| \rangle$ as a function of redshift. The grey band in the Figure shows the redshift range where the simulated volume-averaged ionised hydrogen fraction is $0.1 < f_\mathrm{HII} < 0.9$, approximately spanning the time frame where the ionization fronts are sweeping through the IGM. The latter is the time where the Biermann and Durrive batteries are expected to be the most efficient. Indeed, the \fbbier and the \fbdurr runs show a huge, quick increase in $\langle |\mathbf{B}| \rangle$ shortly after this period of time, consistent with the magnetic fields being generated during the EoR being amplified in galaxies until saturation. Also the \fbcosmo run shows a boost in the volume-averaged magnetic field strength, although slightly delayed, signaling the fact that ionization fronts are responsible for the generation of the magnetic seed, but galactic processes are the source of amplification. Indeed, a similar rapid boost in the volume-averaged strength of magnetic fields was seen at approximately the same redshift in \citet{Marinacci+2015}, where radiation was not followed and reionisation was approximated by a uniform ultraviolet background. Hence, this amplification comes as a consequence of the star-formation and galaxy assembly processes, which reach their peak approximately at the same time \citep[see \eg][]{Madau&Dickinson2014}. 
Finally, the \fbsne run shows a somewhat different evolution, with the volume-averaged magnetic field rapidly increasing after the formation of the first star in the simulation (at $z\approx14$, corresponding to where the solid line in the Figure begins), and peaking at $z\approx2$, when the star-formation activity in the Universe is at its maximum. Interestingly, the volume-averaged magnetic field strength slightly decreases afterwards, because the magnetic fields inside virialised structures have reached saturation already, while outside of them gas keeps expanding, overall reducing the magnetic field strength.

\enrico{Before moving on, we discuss here the coherence length $L_\mathrm{c}$ of the simulated magnetic fields in the \textsc{FB} simulations. We computed this quantity through direct ray tracing, \ie by placing randomly-oriented skewers through the simulation box and computing the path length sharing the same orientation of the parallel magnetic field. \citet{Akahori&Ryu10} showed that this more rigorous approach provides results that are comparable with approximate estimates obtained from the power spectrum of the magnetic field. For our simulation set, at $z=0$ we obtain very similar values of $L_\mathrm{c} \approx 0.33 \, h^{-1}\,\mathrm{Mpc}$ in the \fbbier, \fbdurr, and \fbsne runs, while a larger value of $L_\mathrm{c} = 0.49 \, h^{-1}\,\mathrm{Mpc}$ is found in \fbcosmo, as a consequence of the large-scale coherence imposed in the initial conditions of the latter. We further confirmed that this is the reason by computing $L_\mathrm{c}$ at a few higher redshifts. The \fbcosmo run is the only one showing a \textit{larger} (comoving) coherence length at earlier times, as a consequence of structure formation actually decreasing the field coherence. In the other seeding mechanisms investigated, the field coherence slightly increases with time. These values of $L_\mathrm{c}$ are compatible with measurements of \eg \citet{AlvesBatista&Saveliev20}. }

\subsection{Magnetic fields in the volume-filling phase}

\begin{figure}
    \centering
    \includegraphics[width=\columnwidth]{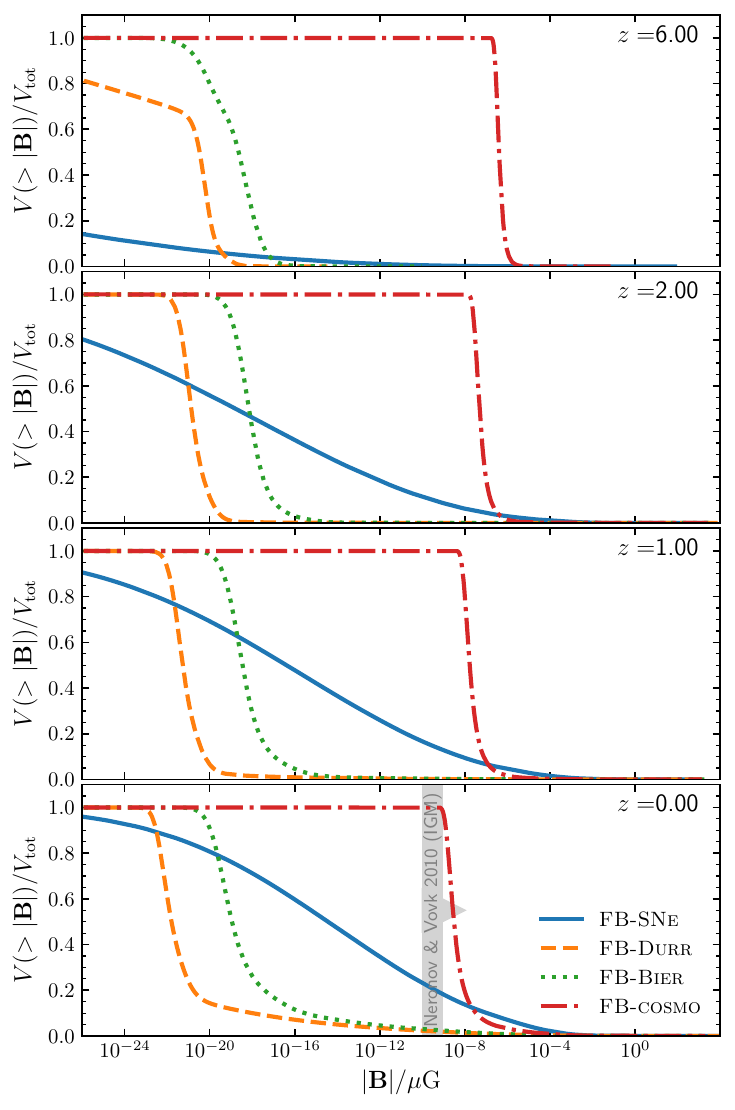}
    \caption{Cumulative volume fraction of magnetised gas at different redshifts in the \textsc{FB} runs. The grey band shows the lower limit on IGM magnetic fields inferred by \citet{Neronov&Vovk2010}.}
    \label{fig:f_magnetised}
\end{figure}

In principle, it is possible to tell apart local and global seeding mechanisms from the spatial distribution of $z\sim0$ magnetic fields, provided the key assumption that the dispersal process of magnetised gas is not efficient \citep{Cho2014}. This argument can be cast into a comparison of the volume filling factors of magnetised gas in the different seeding models. We show this quantity in Fig.~\ref{fig:f_magnetised}, where the cumulative volume fraction magnetised above a certain field strength is shown at four different redshifts (reported in the top right corner of each panel) for the \textsc{FB} runs. It is clear that the \fbcosmo run provides the most magnetisation at all redshifts, simply because the other seeding mechanisms require time not only to generate and amplify magnetic fields, but also to distrbute them outside of overdense regions. Note also that the sharp drop from $V(>|\mathbf{B}|)/V_\mathrm{tot} = 1$ corresponds to the initial seed value rescaled by the expansion of the Universe. It is however important to appreciate that most of the current constraints on the magnetic fields in the Universe come from overdense regions (galaxies and galaxy clusters), which only cover a negligible fraction of the total volume. At $z=6$, in and around these regions, the fields produced by SNe are the strongest among the seeding models investigated (see \eg Fig.~\ref{fig:split_panels_fb} and Sec.~\ref{sec:rm}). In the \fbbier and \fbdurr runs, the distribution shows a knee (most prominent in the latter) at $V(>|\mathbf{B}|)/V_\mathrm{tot} \sim 0.6$. This is a sign of \textit{ongoing} creation of magnetic fields at the ionisation front. Such a feature is less visible in the Biermann battery case since, unlike for the Durrive battery, magnetisation can occur even without ionisation fronts and hence: (i) starts earlier (see also Fig.~\ref{fig:vol_avg}), and (ii) can occur also in regions of space not yet reached by ionising photons.

By $z=2$ all models have magnetised the entire volume, although with very different field strengths. It is interesting to note how the curves corresponding to the \fbcosmo, \fbdurr, and \fbbier runs closely resemble each other, except for being shifted along the horizontal direction. This shows that each of these seeding mechanisms creates a magnetic field which has a relatively homogeneous strength in most of the volume. Conversely, the \fbsne run shows a very different distribution, with regions of large and significantly smaller magnetic strengths coexisting, since the SN injection is a very local process. Interestingly, moving to lower redshift, the magnetic field \textit{decreases} in most of the volume for all models except for \fbsne. This shows that, after the initial magnetisation of the low-density gas, the suppression due to the expansion of the Universe dominates there over amplification mechanisms. We stress that the situation is the opposite at high density, as we discuss \eg in Sec.~\ref{sec:haloes} and Sec.~\ref{sec:rm}.

In the lowest panel of Fig.~\ref{fig:f_magnetised} we show the constraint on IGM magnetic fields from \citet{Neronov&Vovk2010}, whose uncertain reliability needs to be kept in mind \citep[][see discussion above in this section]{Broderick+12}. Again, if taken at face value, this lower limit is fully consistent only with the cosmological seeding mechanisms, while the SN injection model is in agreement only in $\sim 25$\% of the simulation volume. 

Note that the argument laid out above is less affected by resolution limitations in the magnetic field amplification than considerations on the \textit{value} of the magnetic field strength itself. In fact, the volume fraction with significant magnetic fields primarily depends on the ejection of highly-magnetised gas from the galaxies, and its dissemination in the rest of the universe. These are much better converged with resolution than the dynamo-driven magnetic field amplification \citep[see \eg][]{Pakmor+2017}. While a quicker amplification entails a larger gas magnetisation at high redshift, and hence a longer time to distribute the latter in the IGM, Fig.~\ref{fig:f_magnetised} clearly shows that most of the volume is not magnetised by such processed gas.

\subsection{Rotation measures}
\label{sec:rm}

\begin{figure*}
    \centering
    \includegraphics[width=\textwidth]{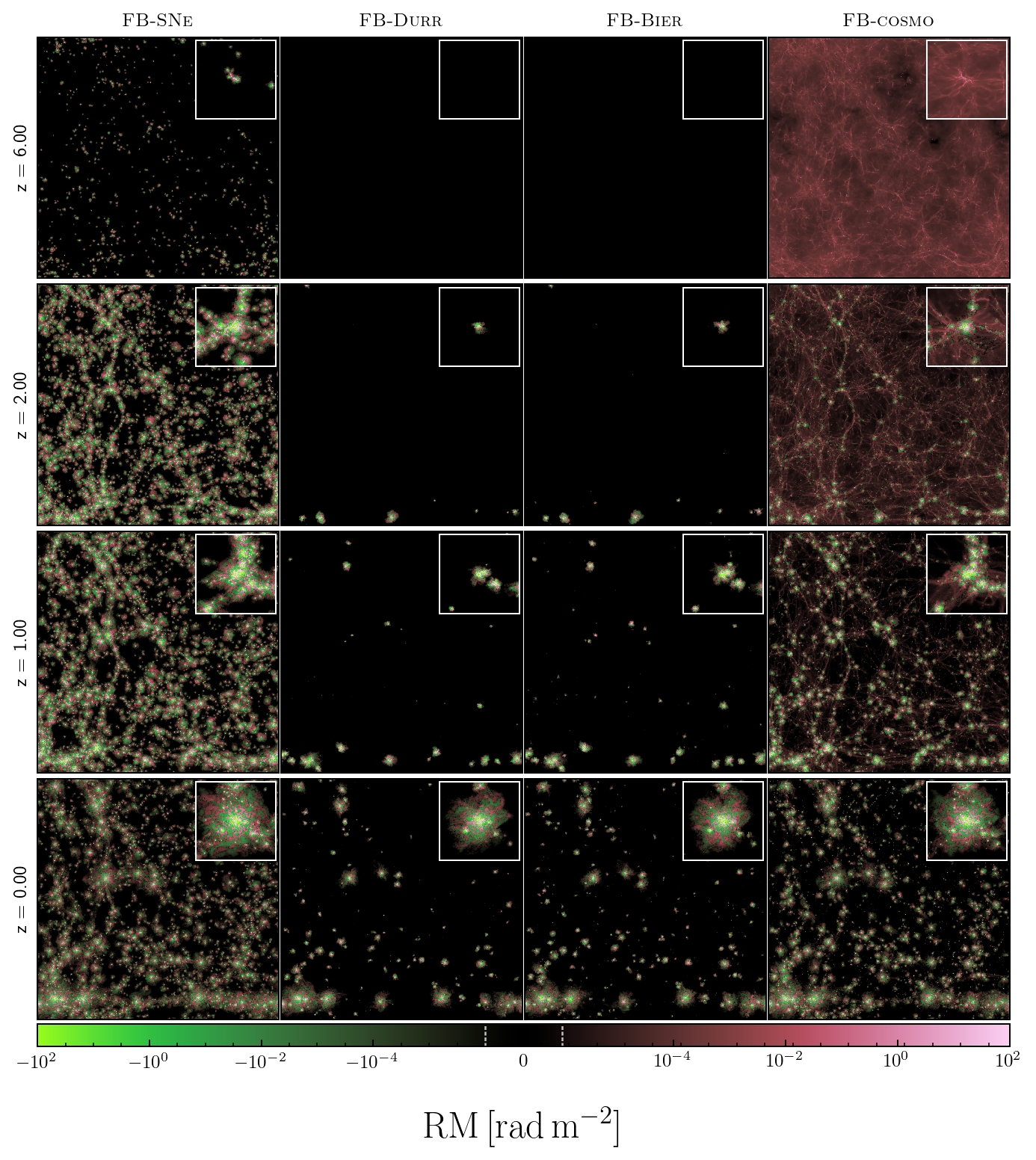}
    \caption{Rotation measure map at different redshift for the \textsc{FB} runs. Each panel shows a projection along the $z$-axis of the entire simulation domain. Note that this is the initial direction of the seed field in the \fbcosmo run. The color scheme is linear in the range $[-10^{-6}, 10^{-6}]$ (marked with dashed lines in the colorbar) and otherwise logarithmic for the magnitude while preserving the sign. The insets in the top right corner of each panel show a zoom-in on the largest halo in the box. Each one of them is obtained as in the larger panels, but using a cube with side $5\,\hMpc$ centred on the halo.}
    \label{fig:rm}
\end{figure*}

While the magnetic field strength at low-redshift appears similar in all models, the magnetic field orientation can be very different. In particular, underdense regions  retain memory of the initial magnetic field when it has a cosmological origin \citep[see \eg][]{Marinacci+2015}. A common way to measure magnetic field strength in diffuse gas is by means of the so-called Faraday rotation measure (RM), \ie through the wavelength-dependent change in the polarisation vector of polarised radiation propagating through a magnetised plasma. It is defined as
\begin{equation}
\label{eq:rm}
    \mathrm{RM} = \frac{e^2}{2 \pi m_\mathrm{e}^2 c^4} \int_0^L n_\mathrm{e} (s) B_\parallel (s) \, \mathrm{d}s ~,
\end{equation}
where $e$, $m_\mathrm{e}$, and $n_\mathrm{e}$ represent the electron charge, mass, and number density, $c$ is the speed of light in vacuum, and $B_\parallel$ is the component of the magnetic field parallel to the line of sight path $\mathrm{d}s$ of total length $L$. For a magnetic field of uniform intensity, the RM also represents a measure of the field coherence length thanks to its integral nature. However, in realistic configurations, this is degenerate with the field strength and electron density distribution. 

We compute simulated RM maps by integrating along the $z$ direction (\ie the direction of the seed field in the \fbcosmo run) for the entire length of the simulation box (\ie $\Lbox = 25 \, \hMpc$), and present 2-dimensional maps of the resulting RM in Fig.~\ref{fig:rm} at $z=6$, $2$, $1$ , and $0$ (top to bottom rows) for the four seeding models we analysed (left to right columns). It appears immediately evident that the different models produce very different RM distributions. At $z=6$, the cosmological seeding produces RMs in the entire projection (although small), with a rather smooth distribution.\footnote{Notice that this is less smooth with respect to other numerical studies \citep[eg][]{Marinacci+2015, Marinacci&Vogelsberger2016, Vazza2017}, since we self-consistently include radiation transport, rather than using a tabulated smooth UV background, which enhances variations in the electron number density.} This is the case because at such high redshifts most of the magnetic fields still preserve their original orientation, hence they coherently contribute to the RM integral. We show the projection along the $z$-axis to render this effect apparent, while projections along different direction show vanishing RM (for the same reason). In the \fbsne run, the relatively strong injected seed fields create small regions of high RM at the location of the first stars. In the other two mechanisms, the magnetic field is instead low and incoherent, hence creating only very small RM.

Moving to lower redshifts, the RM values increase in all models --~although more significantly in the \fbsne and \fbcosmo runs~-- in correspondence with the density peaks, showing once again that the structure formation process is the driver of magnetic field amplification. In the \fbsne run, the highest RM is reached at intermediate redshifts ($z=2$, and $1$), and slightly decreases towards $z=0$ 
(see also the discussion of Fig.~\ref{fig:vol_avg}). In the \fbcosmo run, the RM outside of virialized structures decreases with redshift, because hierarchical structure formation and the expansion of the Universe reduce the gas (and hence electron) density. Finally, the \fbdurr and \fbbier runs show significantly less structures with high RM, in agreement with --~and for the reasons detailed in~-- the discussion of Fig.~\ref{fig:B_vs_mvir_z0}.

\begin{figure}
    \centering
    \includegraphics[width=\columnwidth]{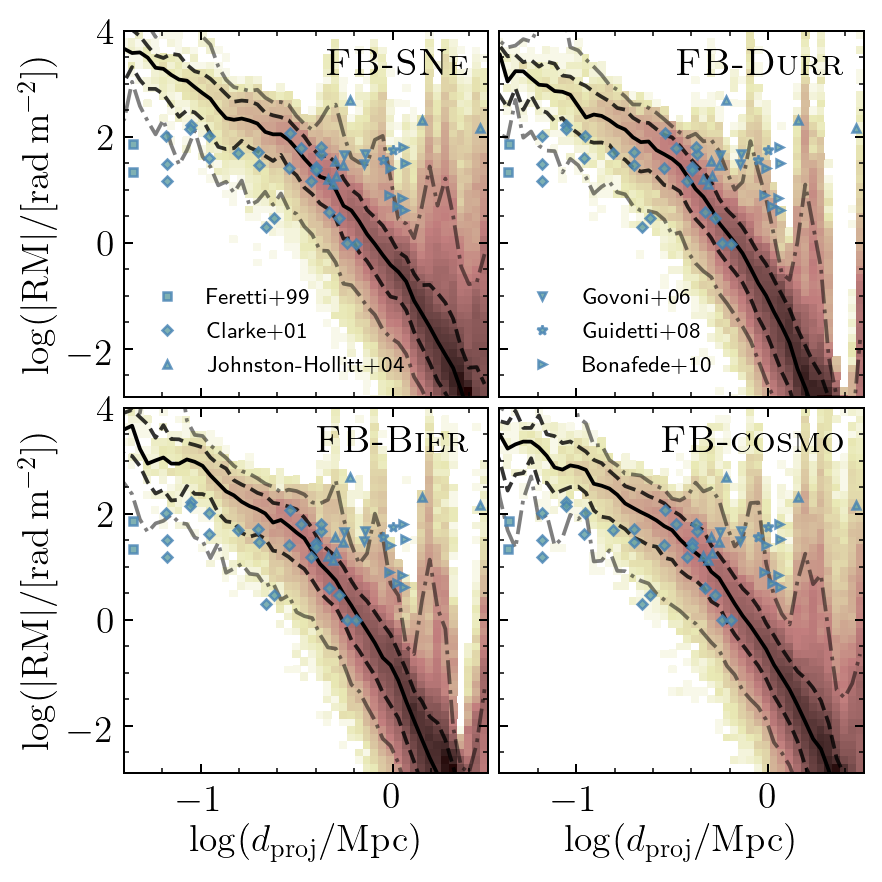}
    \caption{Two-dimensional distributions of rotation measures and projected distance from the largest halo in the \textsc{FB} runs at $z=0$. The color shading intensity scales logarithmically with the number of pixels in each bin. The solid, dashed, and dot-dashed lines show the median, 16th/84th, and 2nd/98th percentiles, respectively. Symbols show measurements taken from \citet{Feretti+99, Clarke+01, Johnston-Hollitt+04, Govoni+06, Guidetti+08, Bonafede+10}.}
    \label{fig:rm_vs_r}
\end{figure}

These differences provide a low-$z$ observational signature of different seeding processes. To determine if current observations are able to effectively distinguish them, we show a zoom-in on the most-massive halo in the simulation (with mass $M_\mathrm{halo} = (3.2 \times 10^{11},\; 1.1 \times 10^{13},\; 2.1 \times 10^{13},\; 8.6 \times 10^{13}) \, \hMsol$ at $z = 6$, $2$, $1$, and $0$ respectively,  virtually identical in all the models) in the top right corner of each panel of Fig.~\ref{fig:rm}. Around these overdensities the RM evolves in a similar way to the rest of the box, although at a faster pace. In particular, by $z=1$ the RM appears very similar in the \fbsne and \fbcosmo runs, and by $z=0$ all models show very similar RM distributions. The only noticeable difference occurs in the outskirts of the halo, where the differences in mildly-overdense gas (see Fig.~\ref{fig:B_vs_rho_z0}) can be appreciated. Quantitatively, we compare our simulations to available observations in Fig.~\ref{fig:rm_vs_r}, where we show the two-dimensional distribution of RM and projected distance from the centre of the largest halo in the simulation (shown as color shading with an intensity that logarithmically scales with the number of pixels in each bin), as well as the median, 16th/84th and 2nd/98th percentiles (solid, dashed, and dot-dashed lines, respectively), and a collection of observations (symbols) from \citet{Feretti+99, Clarke+01, Johnston-Hollitt+04, Govoni+06, Guidetti+08, Bonafede+10}. It should be noted that these observations mostly come from clusters of galaxies which are significantly more massive than the largest halo in our simulation (because of our limited volume). Hence, we expect them to extend to larger radii than in our simulation. For small projected distances, however, our data is only marginally consistent with the available measurements, possibly because of the exclusion of black holes from the galaxy formation model employed, which have the effect of decreasing the central density of massive haloes\enrico{, and hence their RM (through a boosted electron density) at a given magnetic field strength. In fact, the halo magnetic field strength is independent of the halo mass, at least at the massive end (see Fig.~\ref{fig:B_vs_mvir_z0})}. The most relevant feature of Fig.~\ref{fig:rm_vs_r} for the analysis presented in this paper, however, is the fact that different injection mechanisms produce essentially indistinguishable distributions, in line with the results of  Fig.~\ref{fig:B_vs_mvir_z0}. This shows that not only the volume-averaged value but also the radial distribution is similar in all seeding models. 

Many radio galaxies at $z>2$ show RMs in excess of $1000 \, \, \mathrm{rad \, \, m^{-2}}$ \citep[\eg][]{Athreya+98, Pentericci+00,Vernstrom+18}, with the highest value exceeding $15000 \, \, \mathrm{rad \, \, m^{-2}}$ \citep{Broderick+07}. In order to compare our simulations to these data, we produce rotation measure maps identical to those shown in the inset panels of Fig.~\ref{fig:rm} for the $100$ most massive haloes in the \textsc{FB} runs at $z=2$. We then select only pixels within $0.1$ times the halo virial radius. Values of $\mathrm{RM}>1000 \, \, \mathrm{rad \, \, m^{-2}}$ are found in approximately $4$\% of the sightlines, with only a very mild dependence on the seeding process employed (namely, \fbsne produces the largest number of such sightlines --~5\%~-- followed, in that order, by \fbcosmo, \fbdurr, and \fbbier). These numbers, however, strongly depend on the halo mass. Repeating the same analysis for the $10$ most massive haloes boosts the sightline fraction to $\sim20$\%. We conclude that all seeding processes we investigated are consistent with the observed large RMs, and --~hence~-- the latter are unable to constrain any of the seeding mechanisms investigated.

\subsection{Impact on the star formation rate density}
\label{sec:sfrd}

\begin{figure}
    \centering
    \includegraphics[width=\columnwidth]{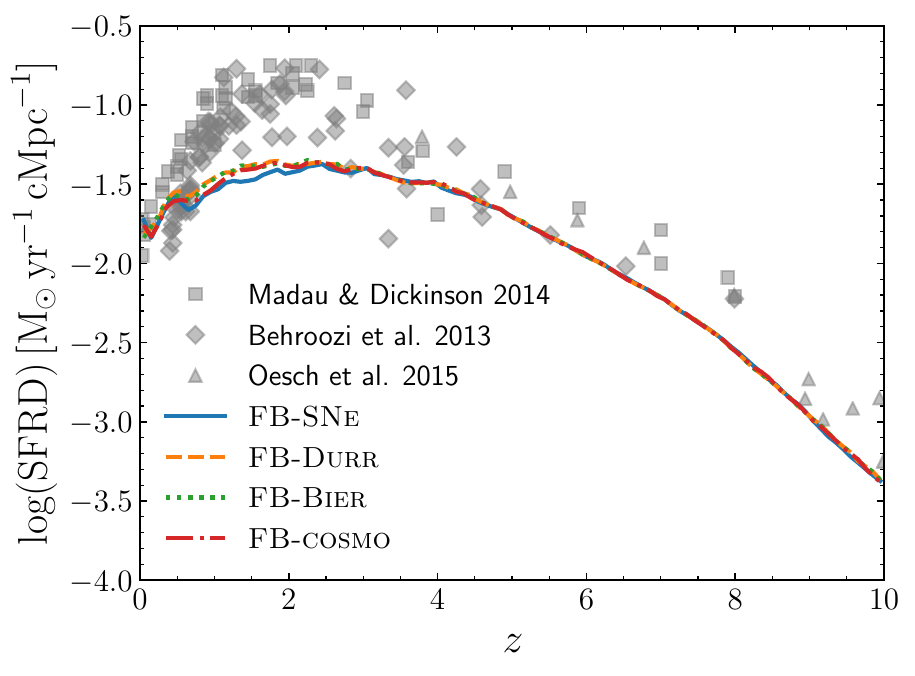}
    \caption{Comoving star-formation-rate density in the \textsc{FB} runs, compared to the collections of data from \citet{Behroozi+2013}, \citet{Madau&Dickinson2014}, and \citet{Oesch+2015}.}
    \label{fig:sfrd}
\end{figure}

Finally, we discuss the impact of magnetic fields on the star formation rate density (SFRD) in the Universe. Magnetic fields influence this quantity through magnetic pressure, which provides additional support against gravitational collapse. For instance, \citet{Marinacci&Vogelsberger2016} showed that for a cosmological seed above a critical value of $B_\mathrm{crit} \sim 10^{-9} \, \mathrm{G}$ in the initial conditions, star formation is suppressed. While in Fig.~\ref{fig:B_vs_rho_z0} we show that at $z=0$ the high-gas-density branch of the $|\mathbf{B}|$ - $\rho_\mathrm{b}$ relation is very similar in all seeding models, and hence the late-times effect of magnetic fields on star formation (occurring in high density regions) should be the same, this relation is established at different times. 

Hence, we explicitly check the effect on the SFRD in the \textsc{FB} runs in Fig.~\ref{fig:sfrd}, where the latter is shown as a function of $z$, together with data collected by \citet{Behroozi+2013}, \citet{Madau&Dickinson2014}, and \citet{Oesch+2015}. The first striking feature is that the SFRD evolution is largely insensitive to the magnetic seeding mechanism. 
Secondly, the 
simulated SFRD falls short of the observed data points at cosmic noon ($z\sim2$), and partially also at higher redshift. The reason for this mismatch is likely to be found in the galaxy formation model employed, which has been created and calibrated to reproduce Milky Way-like galaxies. In fact, a similar version of this model tuned to reproduce cosmological variables was used in \citet{Marinacci&Vogelsberger2016}, and the SFRD evolution was correctly captured. Additionally, we do not include black holes in the simulations. However, they are not expected to boost the formation of stars. We stress that this mismatch with observations does not affects our results since (i) we are interested in \textit{relative} differences between the magnetic seeding models, and (ii) by $z\sim3$ --~when the mismatch becomes significant~-- magnetic fields have already reached their saturation level in most models (see \eg Sec.~\ref{sec:mag_growth}).

\section{Discussion}
\label{sec:discussion}

Our results from  the SN injection scheme are qualitatively similar to those of \citet{Beck+13}, although the latter focuses only on a single Milky Way-like galaxy. For instance, the volume-averaged magnetic field strength (Fig.~\ref{fig:vol_avg} and figures 8 and 11 in their paper) shows both a similar evolution (\ie an initial rapid increase, a peak, and a slow decrease afterwards) and the same relative importance with respect to the magnetic field generated by a cosmological seed. The volume-averaged magnetic field strength presented in figures 8 and 11 of \citet{Beck+13} is systematically larger than ours, but this is just a consequence of their analysis being restricted to a highly-biased region of the universe, namely the halo of a Milky Way-like galaxy. We checked that this is the case by computing the volume-averaged field strength within the main halo in the \textsc{Au} simulations, which yields results similar to the ones reported in \citet{Beck+13}. Similarly, the volume fraction in the same halo has a qualitatively similar evolution to the one reported in figure 10 of \citet{Beck+13}, although at high redshift our simulations have a significantly-larger portion of gas with $|\mathbf{B}| > 10^{-15} \, \mathrm{G}$, potentially because of the different way the seed field is injected from supernovae. By $z\sim1$, however, this difference is erased.

\subsection{Combination of multiple seeding processes}
One limitation of our work is that we do not simulate combinations of multiple seeding mechanisms. Thanks to amplification erasing every memory of the original magnetic seed, the results from simulations with multiple seeding mechanisms can be approximated by combining separate (single-seeding) runs under the rule $\mathbf{B}_\mathrm{S1+S2} (\mathbf{x}) = \max \{\mathbf{B}_\mathrm{S1} (\mathbf{x}), \mathbf{B}_\mathrm{S2} (\mathbf{x}) \}$, where S1 and S2 denote two different seeding schemes, and $\mathbf{x}$ indicates the spatial coordinates. In fact, the strength of amplified magnetic fields is set by structure formation and not by the seeding process. In low density regions, instead, multiple seeding mechanisms will concur to magnetise different regions of the IGM, and hence only the dominant one will set the \textit{combined} magnetic field strength at any given point in space. Using this simple combination rule, we can 
predict that combining SN injection and cosmological seeds will produce a configuration where the latter dominates in the low-density gas covering most of the volume, while a combination of the two is responsible for magnetic fields in collapsed structures like haloes. This is exactly the conclusion reached in \citet{Katz+19}, where a tracer scheme for magnetic fields was used to study the contribution of cosmological and SN-injected fields down to $z=6$. Hence, we feel confident that our simple summation rule can provide some insights on the results of combining multiple seeding schemes.

The Biermann and Durrive batteries are very similar in their behaviour. However, the latter is systematically lower in strength (by approximately $3$ orders of magnitude) before amplification. This is not surprising, as they are powered (at least on large scales) by the same physical process, \ie ionisation fronts. In fact, the latter not only generate an electric field (thanks to charge segregation during photo-ionisation) perpendicular to them, but also a temperature gradient along the same direction. Hence, in realistic configurations, both the Durrive and the Biermann battery terms will be at play. Using the simple addition rule above, it is easy to see that 
in virtually all realistic configurations, the Durrive mechanisms will be subdominant, rendering the Biermann battery entirely sufficient to capture large-scales magnetisation effects due to plasma processes during galaxy formation. While there is the theoretical possibility of having no Biermann battery but a non-zero Durrive seed (thanks to the second term on the right-hand-side of Eq.~\ref{eq:DurriveBattery}), in practice this is extremely unlikely since it requires perfect alignment of density and temperature gradients.

\subsection{Considerations concerning galaxy formation simulations}
Another interesting outcome of our study is that, for what concerns the properties of galaxies at low redshift, the different magnetic seeding mechanisms do not have any discernible impact, as long as saturation is reached. Based on this result, we believe it is worthwhile for future galaxy formation simulations to consider the inclusion of physically motivated seeding prescriptions such as the Biermann battery instead of making ad-hoc assumptions about a pre-existing  primordial seed field. This will increase the physical fidelity of the numerical models without requiring any additional free parameters (for \eg the Biermann battery, since all physical quantities involved are self-consistently followed in the simulation itself). In contrast, the commonly employed cosmological seeding requires arbitrary choices regarding the initial field strength and geometry, which can even leave an imprint in the field configuration in low-density regions \citep[see \eg][]{Marinacci+2015}. At a practical level, this replacement is unlikely to require any re-calibration of galaxy-formation models in the majority of cases, as they are typically tuned against low-redshift observations, where magnetic field properties are not affected by the seeding mechanism. Nevertheless, the different seeding mechanisms may have an unequal impact on $z \gtrsim 2$ galaxies, if the amplification mechanisms are rapid enough in proto-galaxies to provide a significant amount of magnetic pressure already in high-redshift star-forming clouds. However, tailored simulations are necessary to further investigate this possibility.

Ongoing and forthcoming surveys of Fast Radio Bursts (FRBs) will likely provide growing samples of RMs. Eventually, the redshift of detected FRB sources will become sufficiently large to seriously test different magnetic seed field models \enrico{\citep[see \eg][]{Akahori+14a,Akahori+14b,Hackstein+19,Hackstein+20}}. For instance, the first phase of the Square Kilometre Array is expected to provide spectro-polarimetric images of approximately $10^4$ pulsars and $10^7$ compact extragalactic sources \citep{Beck07}.
In the meantime, simulations need to strive to achieve sufficient resolution to reliably simulate the amplification of magnetic fields on galactic and meta-galactic scales in a quantitatively robust way.

\section{Conclusions}
\label{sec:conclusions}

In this paper, we have analysed the production, amplification, and dispersal of magnetic fields in four different magnetic seeding models, namely: a pre-recombination uniform seed, injection from supernova explosions, the Biermann battery, and the recently-proposed Durrive battery at play during the Epoch of Reionization. Our study was carried out using two sets of numerical radiation-magneto-hydrodynamical simulations, one employing the zoom-in technique to reach high resolution in a single galaxy, and one distributing the numerical resources equally on a large patch of the Universe. Both sets include the same state-of-the-art galaxy formation model.
Our main results can be summarised as follows:
\begin{itemize}
    \item In all seeding mechanisms, the simulated magnetic fields are virtually identical both within galaxies (Fig.~\ref{fig:B_center_evol_Auriga}) and haloes (Fig.~\ref{fig:B_vs_mvir_z0}). However, the timing of when this state of saturation is reached is very different. 
    
    \item For all seeding schemes, there exists a spatial correlation between metallicity and magnetic field strength. The time when this correlation emerges is different in different schemes, and depends on the seed field strength and its injection time.
    
    \item In all models and at all redshifts, the magnetic field impact on galaxy evolution is similar (Fig.~\ref{fig:histogram_beta_inv_Auriga} and Table~\ref{table:gal_prop}).
    
    \item There exists a halo mass threshold above which the halo magnetic field is completely insensitive to the seeding mechanism (Fig.~\ref{fig:B_vs_mvir_z0}). This threshold is slightly different for each seeding scheme, and depends on a combination of seed field strength and injection time. Below the threshold, halo magnetic fields reflect the ones in the gas they accrete.
    
    \item In all seeding models, except for the SN injection, underdense and mildly-overdense gas follows a density--magnetic strength relation (Fig.~\ref{fig:B_vs_rho_z0}), which is consistent with pure flux conservation for the Durrive battery and the cosmological seeding, while being slightly steeper for the Biermann battery. 
    
    \item At gas overdensity of $\sim 100$ the magnetic amplification mechanisms become dominant and establish a new density--magnetic strength relation offset from the original one, due to compression/expansion of highly-magnetised gas (Fig.~\ref{fig:B_vs_rho_z0}).
    
    \item The volume filling fraction as a function of magnetic field strength is very different in the models we investigated (Fig.~\ref{fig:f_magnetised}). In the SN injection scheme, regions of very strong and very weak field co-exist, while other seeding mechanisms impose a floor on the magnetic field strength in the Universe. 
    
    \item Rotation measure maps (Fig.~\ref{fig:rm}) capture the differences among the models in the highly magnetised gas. However, in the proximity of massive objects (where most of the available observation lie), all seeding processes produce very similar results, which agree relatively well with observations (Fig.~\ref{fig:rm_vs_r}).
    
    \item The impact of the magnetic seeding mechanisms on the star-formation rate density of the universe is generally negligible (Fig.~\ref{fig:sfrd}).
    
\end{itemize}

In conclusion, all the seeding models investigated produce indistinguishable magnetic field configurations at $z\sim0$ in the regime where observations are available (with the possible exception of the lower limit reported in  \citenp{Neronov&Vovk2010}, which is however heavily debated, see \eg \citenp{Broderick+12} and subsequent work). We have uncovered differences in the magnetisation of the Universe produced by these different seeding schemes, which can provide precious information for future attempts at understanding the creation of magnetic fields in the Universe. We have shown for the first time that the seeding scheme recently proposed in \citet{Durrive&Langer2015, Durrive+2017} can magnetise the Universe in realistic scenarios. However, our results show that the Biermann battery produces geometrically similar but stronger magnetic seed fields, hence being almost always dominant over the Durrive battery.

Although the origin of magnetic fields in the Universe remains an undecided mystery, our results provide an instructive overview of the impacts, similarities, and differences of four magnetic seeding mechanisms. In combination with future observations (\eg from SKA), our results can be a valuable aid in the quest of understanding cosmic magnetism. Notably, our results highlight the importance of reionization-epoch galaxies as a testing ground for the different magnetic field creation mechanisms.

\section*{Acknowledgements}
\enrico{We thank the referee, Franco Vazza, for his insightful comments, which improved the quality of the manuscript. }
We are thankful to the community developing and maintaining software packages extensively used in our work, namely: matplotlib \citep{matplotlib}, numpy \citep{numpy}, scipy \citep{scipy}, cmasher \citep{cmasher}.

\section*{Data availability}
The data underlying this article will be shared on reasonable request to the corresponding author.

\bibliographystyle{mnras}
\bibliography{bibliography}

\appendix

\section{Durrive battery implementation}
\label{app:Durrive_implementation}

\begin{figure}
    \centering
    \includegraphics[width=\columnwidth]{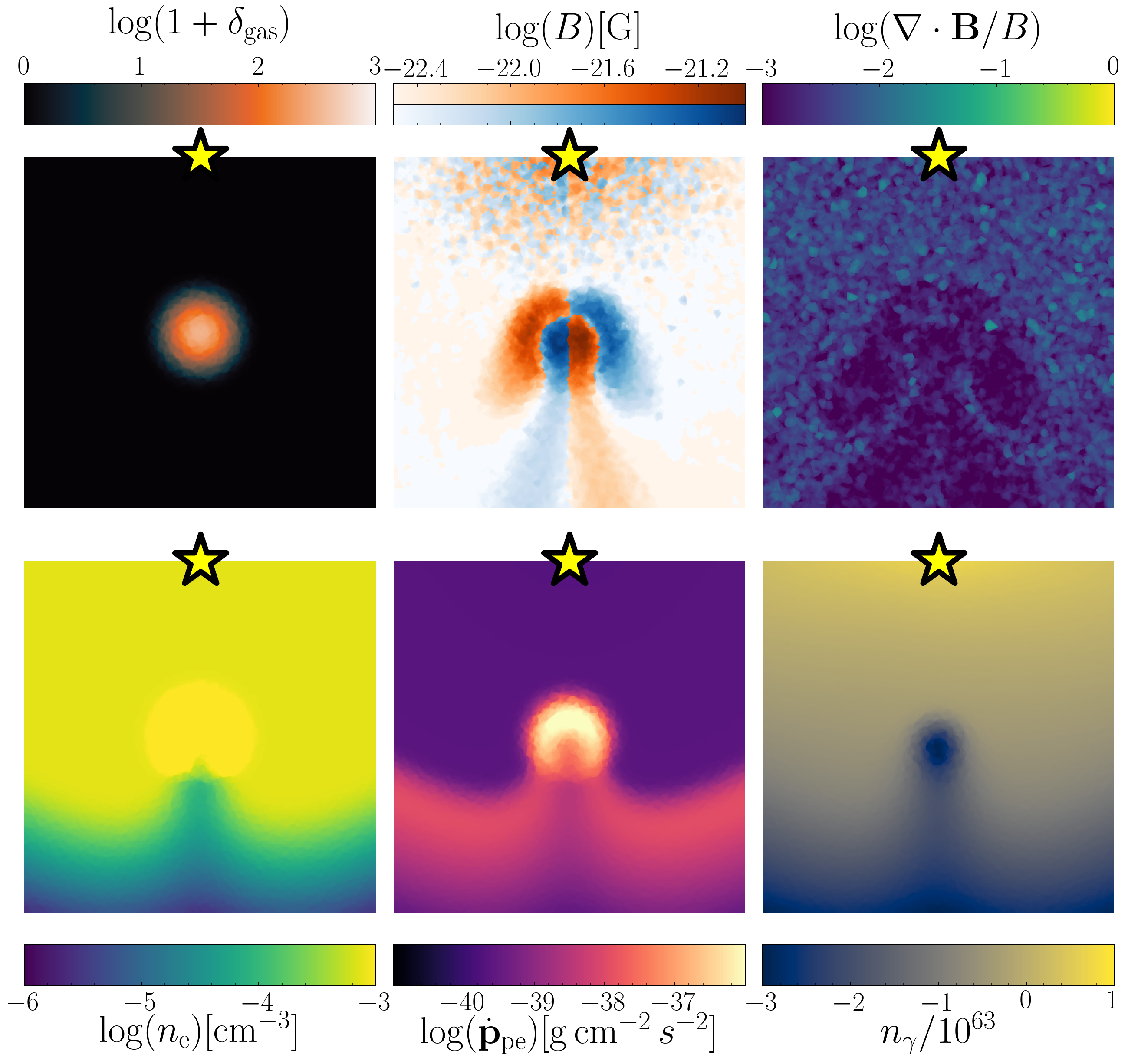}
    \caption{Example of a self-consistent simulation of the Durrive battery mechanism in an idealised setup, \ie a Gaussian overdensity in a homogeneous medium. The panels show six different physical quantities in a thin slice across the center of the simulation box, namely the gas overdensity (top left), magnetic field (top central), normalised field divergence (top right), electron density (bottom left), momentum transferred during photo-ionization (bottom center) and frequency-integrated photon number density (bottom right). The star symbol indicates the location of the photon source.}
    \label{fig:Durrive_implementation}
\end{figure}

An example of the self-consistent calculation described in Sec.~\ref{sec:Durrive_battery} is presented in Fig.~\ref{fig:Durrive_implementation} for an idealised setup. In a simulation box of linear size $\Lbox = 8 \, \hkpc$, we place a single source (indicated by the star symbol) at the centre of one of its faces. The source emits photons at a constant luminosity, following a power-law spectrum with slope $\alpha_\mathrm{QSO} = -2$ (roughly mimicking a quasar spectrum), normalised to a luminosity $L_\nu (13.6 \mathrm{eV}) = 3.313 \times 10^{22} \, \mathrm{erg \, s^{-1} \, Hz^{-1}}$ (corresponding to $5 \times 10^{48} \, \mathrm{ph \, s^{-1}}$ at 13.6 eV). The box contains only gas with constant density and a single Gaussian overdensity at its centre (top left panel). For this test run, we disable hydrodynamics, forcing the density distribution to remain constant. The generated magnetic field (top central panel) closely resembles the analytically predicted one \citep[see \eg][]{Durrive&Langer2015}. While close to the source numerical noise can seed the creation of magnetic fields that can have a similar strength, the associated divergence (top right panel) is significantly higher. Hence, the divergence cleaning algorithm present in \arepo will suppress this field, while leaving the ones produced by the Durrive battery unaltered. The bottom row in the figure shows the relevant physical quantities that enter in the magnetic field creation, namely (from left to right) the electron density, the momentum transferred to the photo-ionised electron, and the photon density (integrated over the entire frequency range covered by the simulation). Note that, for display purposes, the simulation presented in Fig.~\ref{fig:Durrive_implementation} has higher resolution than the cosmological simulations analysed in this paper. However, we have checked that degrading its resolution to match the one of our cosmological simulation set does not alter the outcome.

\section{Supernova magnetic seed strength}
\label{app:compare_sn}

\begin{figure}
    \centering
    \includegraphics[width=\columnwidth]{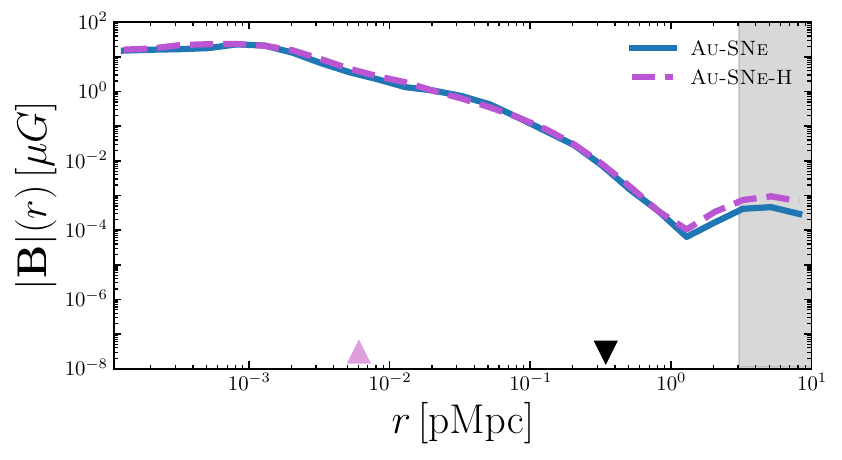} \\
    \includegraphics[width=\columnwidth]{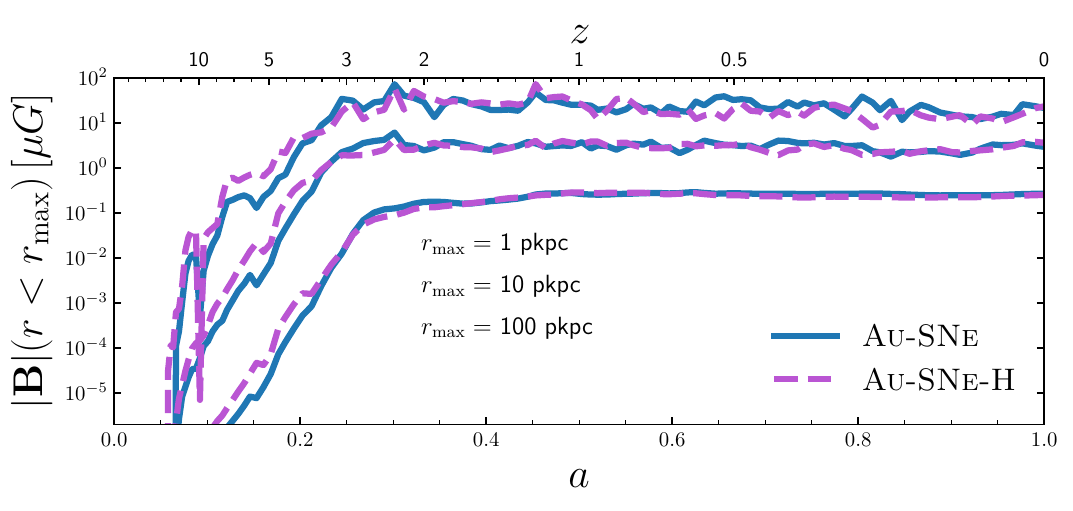} \\
    \includegraphics[width=\columnwidth]{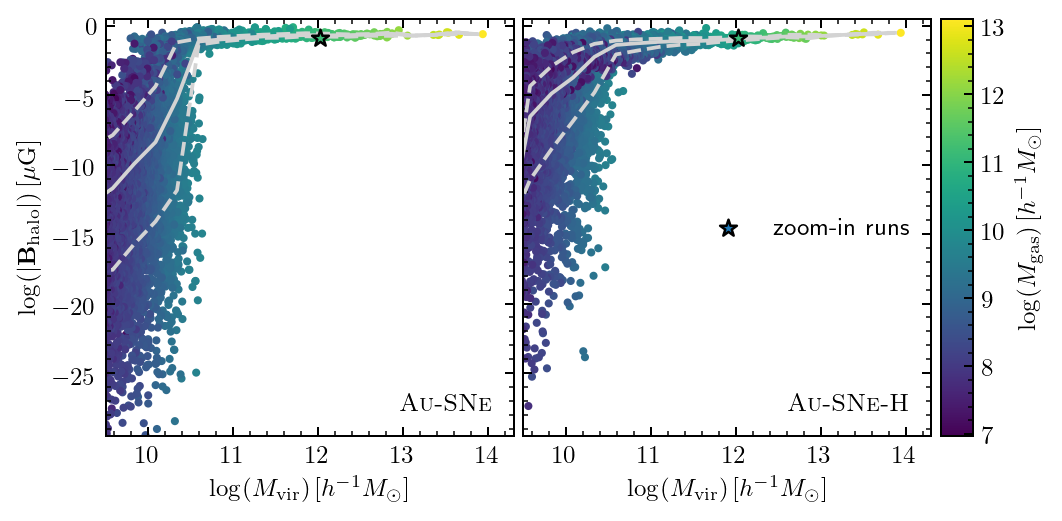} \\
    \caption{Plots comparing the effect on galaxies of two different values for the fraction of SN energy injected into magnetic fields, namely the fiducial $f_B = 10^{-4}$ for \fbsne, and $f_B = 10^{-3}$ for \fbsneH. Each panel reflects exactly one of the Figures in this paper. From top to bottom: bottom right panel of Fig.~\ref{fig:dens_Bfield_Auriga}, Fig.~\ref{fig:B_center_evol_Auriga}, and Fig.~\ref{fig:B_vs_mvir_z0}.}
    \label{fig:SN_comparison_galaxies}
\end{figure}

\begin{figure}
    \centering
    \includegraphics[width=\columnwidth]{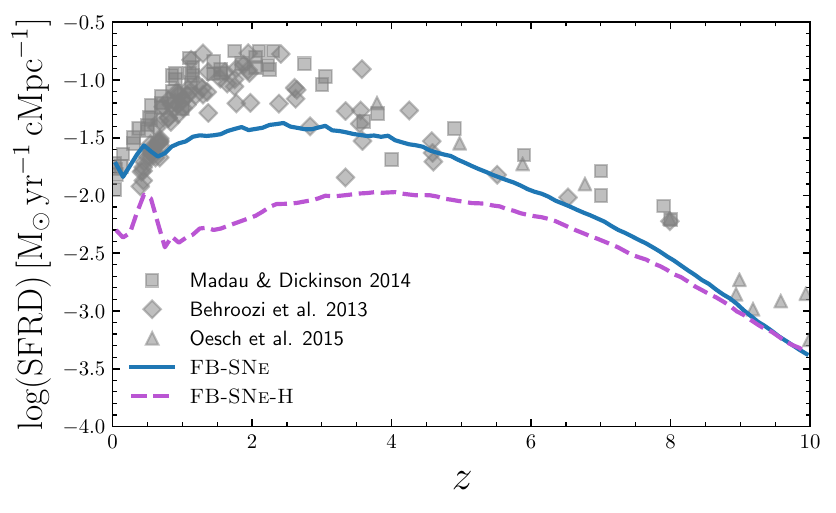} \\ 
    \includegraphics[width=\columnwidth]{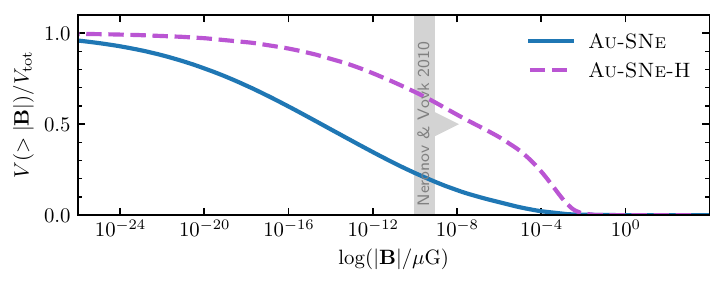} \\ 
    \includegraphics[width=\columnwidth]{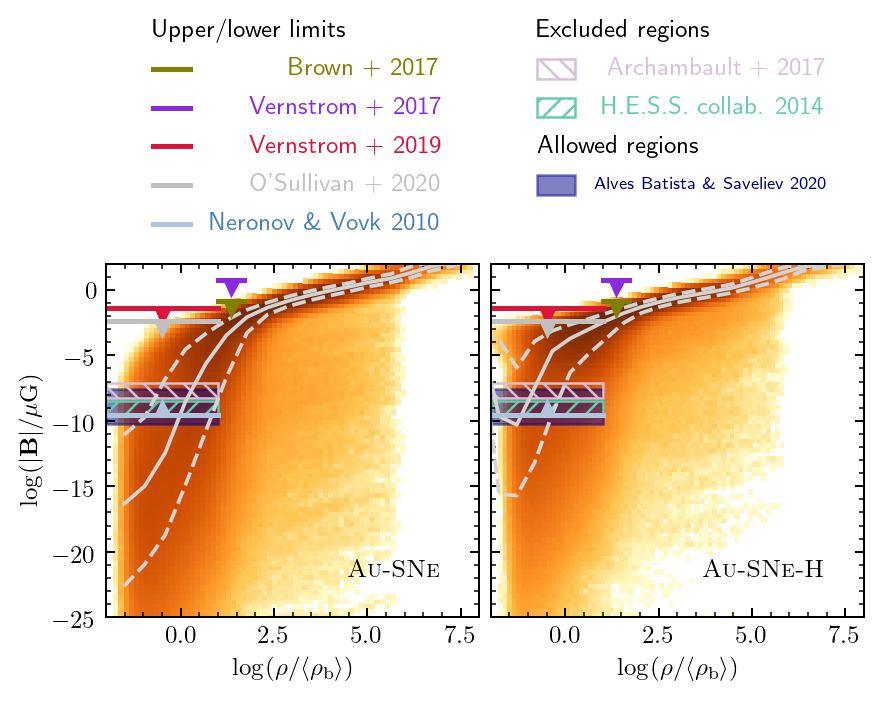} \\
    \caption{Plots comparing the effect on large-scale properties of the universe of two different values for the fraction of SN energy injected into magnetic fields, namely the fiducial $f_B = 10^{-4}$ for \fbsne, and $f_B = 10^{-3}$ for \fbsneH. Each panel reflects exactly one of the Figures in this paper. From top to bottom: the bottom panel of Fig.~\ref{fig:B_vs_rho_z0}, Fig.~\ref{fig:sfrd}, and Fig.~\ref{fig:f_magnetised}.}
    \label{fig:SN_comparison_universe}
\end{figure}

The amount of energy (if any) injected into magnetic fields by each supernova event is unknown. In this Appendix, we briefly explore the impact that this parameter has on galactic and magnetic field properties. To this end, we compare our fiducial value of $f_B = 10^{-4}$ and a value ten times larger. These simulations are labelled \ausne and \ausneH for the Auriga re-simulation, and \fbsne and \fbsneH for the uniform resolution runs, respectively. For our standard supernova energy of $E_\mathrm{SN} \approx 1.69 \times 10^{51} \, \mathrm{erg}$, these values correspond to $E_\mathrm{B,SN} \sim 10^{47}$ -- $10^{48} \, \mathrm{erg}$. These are, for instance, approximately one order of magnitude smaller than those employed in \citet{Katz+19} for a similar numerical experiment. 

We start by comparing the effects on galaxies. As shown in Table~\ref{table:gal_prop}, the galactic properties are only mildly affected by the seed field strength. The largest relative difference is found in the stellar half-mass radius, which is approximately $20$ per cent larger in the run with higher seed field strength. This can be understood as a consequence of the additional magnetic pressure at early times that hinders gas collapse in the innermost regions of the galaxy, hence producing a smoother stellar component. Interestingly, however, the total stellar mass within this radius is not affected by the increased seed field, indicating that gas is not entirely prevented from being accreted, but rather is just redistributed to a larger radius by the increased pressure. 

In Fig.~\ref{fig:SN_comparison_galaxies} we re-create a subset of the figures presented in the main body of the paper, but now comparing runs with different values of $f_B$. The panels show that the value of this parameter does not impact in any way the $z=0$ magnetic field strength in and around galaxies (top panel), and only very mildly the amplification time (middle panel), as can be seen by the fact that the dashed lines are systematically higher than the solid lines until saturation. Finally, the bottom panel of the figure shows that the value of $f_B$ affects the halo magnetic field only for low-mass objects, which in our simulations are tracing the initial seed field injected into the gas (see Sec.~\ref{sec:haloes}).

In Fig.~\ref{fig:SN_comparison_universe} we show plots belonging to the main body of this paper and concerning the largest scales, in a version where different values of $f_B$ are compared. In the top panel, the SFRD evolution is reported. In this case, there is a very prominent difference between the two runs. In fact, when a larger value of $f_B$ is used, the star formation is strongly suppressed and becomes inconsistent with observations. This can be understood as an effect of magnetic pressure preventing the collapse of gas into stars. This is consistent with the result presented in the bottom panel of Fig.~\ref{fig:SN_comparison_galaxies}, although they refer only to $z=0$, showing that low-mass haloes have significantly larger volume-averaged magnetic fields in the \ausneH run. However, since these haloes are close to our resolution limit, the suppression of star formation may be over-estimated for them. For instance, in the simulations of \citet{Katz+19} a magnetic energy approximately $10$ times larger than ours was injected from SN explosions, but such a dramatic effect on the SFRD seemed not to be present, potentially because of their better resolution. However, their simulations stop at $z\sim6$, where the difference between the two values of $f_B$ is still small also in our simulations. Hence, their simulations may simply not capture this effect because they were not run for long enough. In the middle and bottom panels of Fig.~\ref{fig:SN_comparison_universe} we show the effect of increasing $f_B$ on the volume-filling phase of cosmic baryons. The magnetic field in underdense and mildly-dense gas is boosted (rendering this model fully consistent with the \citenp{Neronov&Vovk2010} lower limits), hence increasing the volume filling factor of significantly magnetised gas.

\bsp	
\label{lastpage}
\end{document}